\documentclass[aps,superscriptaddress,eqsecnum,nofootinbib,preprintnumbers]{revtex4}

\usepackage{graphicx,epsfig}
\usepackage{amsmath}
\usepackage {amssymb}
\usepackage{color}
\usepackage[lofdepth,lotdepth]{subfig}

\newcommand{\be}{\begin{eqnarray}}
\newcommand{\ee}{\end{eqnarray}}
\newcommand{\bea}{\begin{eqnarray}}

\newcommand{\eea}{\end{eqnarray}}

\def\a{\alpha}

\def\m{\mu}
\def\n{\nu}

\newcommand{\pkn}{\partial_\nu}
\newcommand{\pkm}{\partial_\mu}
\newcommand{\bma}{\begin{pmatrix}}
\newcommand{\ema}{\end{pmatrix}}
\newcommand{\beno}{\begin{equation*}}
\newcommand{\eeno}{\end{equation*}}

\newcommand{\pa}{\partial}

{}

\begin{document}


\title{Gravitational Collapse of a Homogeneous Scalar Field Coupled Kinematically to Einstein Tensor}

%
\vspace{1cm}

%

\author{George Koutsoumbas}
\email{kutsubas@central.ntua.gr} \affiliation{Department of
Physics, National Technical University of Athens, Zografou Campus, Athens
GR 157 73, Greece}
\author{Konstantinos Ntrekis}
\email{drekosk@central.ntua.gr} \affiliation{Department of
Physics, National Technical University of Athens, Zografou Campus, Athens
GR 157 73, Greece}
\author{Eleftherios Papantonopoulos}
\email{lpapa@central.ntua.gr} \affiliation{Department of Physics,
National Technical University of Athens, Zografou Campus Athens GR 157
73,  Greece}
\author{Minas Tsoukalas}
\email{minasts@central.ntua.gr}
\affiliation{Physics Department, Bo\u{g}azi\c{c}i University, \\
34342, Bebek, Istanbul, Turkey}\affiliation{
  Centro de Estudios Cient\' ificos, Casilla
1469, Valdivia, Chile }\affiliation{Department of Physics,
National Technical University of Athens, Zografou Campus GR 157
73, Athens, Greece}


 \vspace{.2in}


%
\vspace{1.5cm}

\begin{abstract}
We study the gravitational collapse of a  homogeneous
time-dependent scalar field that, besides its  coupling to
curvature, it is also kinematically coupled to the Einstein tensor.
This coupling is a part of the Horndeski theory and we investigate its effect on the collapsing process. We find that the  time required for the scalar field to collapse  depends on  the value of the derivative coupling and the singularity is protected by a horizon. Matching the internal solution with an external Schwarzschild-AdS metric we show that a black hole is formed, while the weak energy condition is satisfied during the collapsing process. The scalar field takes on a finite value at the singularity.
\end{abstract}

\vspace{8.5cm}




\maketitle



\section{Introduction}

Studying the gravitational collapse of matter to form a neutron
star core, Oppenheimer and Volkov \cite{original1} and Tolman
\cite{original2} went beyond Newtonian gravitational theory and
suggested that because of high masses and densities general
relativistic effects must be considered. They also proposed that,
because of the large mass of the neutron star, a
non-relativistically degenerate equation of state might be
more appropriate and most importantly  the gravitational effect of
the kinetic energy of the collapsing matter could not be
neglected.  They based their discussion on a general relativistic
treatment of the equilibrium of spherically symmetric
distributions of matter. In a subsequent paper Oppenheimer and
Snyder \cite{original3} studied the non-static solutions resulting
from the collapse of a spherically symmetric shell  in comoving
coordinates.

Since these pioneering works there has been a lot of activity in
studying the gravitational collapse of a spherical mass shell
and its stability. The strategy was to consider  a fluid with some
kind of symmetry and then solve the resulting Einstein equations
provided that an equation of state is given (for a recent review
see \cite{Joshi:2012mk}). These studies have the disadvantage that
they do not capture the dynamics of the collapsing matter. To
remedy this, the collapse of matter parametrized by a scalar field
has been studied. In this way the dynamics of collapsing matter has been
taken into consideration through the scalar field equation. The
gravitational collapse of scalar fields in classical General
Relativity has been widely studied in the literature. Since the early
 90's,  models of scalar field collapse exhibiting a naked
singularity have been found numerically by Choptuik
\cite{Choptuik:1992jv} and analytically by Christodoulou
\cite{Christodoulou:1984hg, Christodoulou:1994hg, Christodoulou:1999hg}. These models were violating the
so--called Penrose Cosmic Censorship Conjecture
\cite{Penrose:1969pc}. In these  works the scalar field is
massless  and free of any self-interactions.  A class of
potentials was found in \cite{Hertog:2003xg}, where smooth
initial data evolve to give rise to a naked singularity, and
energy conditions may be violated. The role of naked singularities
in gravitational lensing  was studied in \cite{Virbhadra:2002ju}.

Collapsing models with homogeneous scalar fields have been discussed,
exhibiting a  class of potentials determining singularity
formation. A crucial role in determining the causal structure is
played by the existence, or not, of an apparent horizon during the
evolution. The collapsing process of a scalar field to a singularity is quite different from what happens in many examples of matter
models exhibiting a central naked singularity.
  Here, instead, it
may be that the singularity located at the boundary of the $``$ball"
of the scalar field can be naked. Therefore, the scalar field solution
must be matched with a suitable external solution and the
behaviour of the radial geodesic in the external solution must be
studied accordingly.

Conditions under which gravity coupled to self-interacting scalar
field yields a singularity formation were found and discussed in
\cite{Giambo:2005se}. It has been shown that the formation of
singularities of the gravitational collapse of homogeneous scalar
fields with a potential is completely determined by a condition of
integrability of a function related to the energy density of the
model. If this function is bounded, apparent horizon formation is
impossible during the evolution and a naked singularity is formed.
A class of collapsing scalar field models with a nonzero potential has been constructed in \cite{Goswami:2004ne}; the collapse end state for these models was a naked singularity. The weak energy
condition was satisfied by the collapsing configuration. It was
shown that the formation of either a black hole or a naked
singularity as the final state of the dynamical evolution is
governed by the collapse rate. It has been seen that the cosmic
censorship was violated in dynamical scalar field collapse.

The effects of a non-vanishing cosmological term on the final fate
of a spherical homogeneous or inhomogeneous collapsing dust cloud
have been discussed in
\cite{Cissoko:1998mx,Deshingkar:2000hd,Lake:2000rm,Madhav:2005kg}. It was
shown that, depending on the initial situation out of which
the collapse evolves, the result of collapse is the formation of
either a black hole or a naked singularity as the end state of
collapse \cite{Deshingkar:2000hd}. In \cite{Cissoko:1998mx} it was
shown that the cosmological constant term slows down the collapse
of matter, also limiting the size of the black hole.  The
formation of black holes or naked singularities is studied in
\cite{Baier:2014ita} in a model in which a homogeneous
time-dependent scalar field with an exponential potential couples
to a four-dimensional gravity with a negative cosmological
constant. In \cite{Zhang:2015dwu} the collapse of a charged scalar
field in a de Sitter space was studied.

More general scalar field configurations appear in extensions of the standard Gravity in scalar-tensor  theories. One of these scalar-tensor theories
that is extensively studied is the Horndeski theory \cite{Horndeski:1974wa}. This  theory, which has been
recently rediscovered \cite{Deffayet:2011gz} and written in a
simpler form {\footnote{See also extensions of Horndeski theory in
\cite{Gleyzes:2014dya,Gao:2014soa,Padilla:2013jza,Ohashi:2015fma}.}},
 gives second-order field equations and contains as a subest a theory
that preserves classical Galilean symmetry \cite{Nicolis:2008in,
Deffayet:2009wt, Deffayet:2009mn,Deffayet:2013lga}. The action of the theory is
\be\label{horn} S=\int d^{4}x
\sqrt{-g}\Big(\mathcal{L}_{2}+\mathcal{L}_{3}+\mathcal{L}_{4}+\mathcal{L}_{5}
\Big)~, \ee where \bea
\mathcal{L}_{2}&=&G_{2}(\phi,X)~,\\
\mathcal{L}_{3}&=&-G_{3}(\phi,X) \square \phi~,\\
\mathcal{L}_{4}&=&G_{4}(\phi,X)R+G_{4X}(\phi,X)\left[(\square \phi)^{2}-\nabla_{\m}\nabla_{\n}\phi\nabla^{\m}\nabla^{\n}\phi \right]~,\\
\mathcal{L}_{5}&=&G_{5}(\phi,X)G_{\m\n}\nabla^{\m}\nabla^{\n}\phi-\frac{1}{6}G_{5X}(\phi,X)\left[(\square
\phi)^{3}+
\nabla_{\m}\nabla_{\n}\phi\nabla^{\n}\nabla^{\a}\phi\nabla^{\m}\nabla_{\a}\phi-3\square
\phi \nabla_{\m} \nabla_{\n}\phi\nabla^{\m}\nabla^{\n}\phi
\right]~. \eea The functions $G_{i}$, with $i=2,3,4,5$, depend on
the scalar field $\phi$ and its kinetic energy
$X=-\frac{1}{2}\nabla^{\a}\phi\nabla_{\a}\phi$ and $G_{iX}$
denotes just the partial derivative of $G_{i}$ with respect to
$X$, $G_{iX}=\frac{\partial G_{i}}{\partial X}$.

One of the terms in the above action, which has been widely studied,  is the derivative coupling of the scalar field to the Einstein tensor appearing in
the Horndeski Lagrangian 
\be\label{EGBscalaraction1}
 I=\int  d^4x\sqrt{-g}\left[ \frac{R-2\Lambda}{16\pi G}-\frac{1}{2} \left(g^{\mu\nu}-\lambda
 G^{\mu\nu}\right)\nabla_\mu\phi \nabla_\nu\phi  \right]~,\ee
which also includes the  canonical kinetic term. We note here that the term $\lambda G^{\mu\nu}\nabla_\mu\phi \nabla_\nu\phi$ couples the scalar field  directly to the curvature, unlike the remaining terms in the Horndeski Lagrangian (\ref{horn}), which involve higher derivatives of the scalar field.

Local solutions for the above action including the
non-minimal kinetic term with a constant coefficient $G$  have been discussed in several recent papers \cite{Rinaldi:2012vy} evading the non-hair theorem for the Horndeski theory \cite{Hui:2012qt}.  Spherically symmetric black
hole solutions that are asymptotically anti-de Sitter have been found.
In these solutions the scalar field usually diverges on the
horizon. To circumvent the problem of regularity of local
solutions one can break the shift symmetry of the scalar field by
introducing a mass term for the scalar field
\cite{Kolyvaris:2011fk,Kolyvaris:2013zfa}; in this case
spherically symmetric black hole solutions have been found.
Another way to remedy this problem, while keeping the shift
symmetry, is to introduce an additional, mild, linear dependence
in the time coordinate for the scalar field
\cite{Babichev:2013cya,Charmousis:2014zaa}, and stability issues
of this solution were studied in \cite{Cisterna:2015uya,Ogawa:2015pea}. See also
extensions to higher dimensions \cite{Charmousis:2015txa}. More recently there has been some interest in studying the formation of neutron stars and compact objects in Horndeski theory
\cite{Cisterna:2015yla,Silva:2016smx,Maselli:2016gxk,Brihaye:2016lin}. The case of  slowly rotating neutron stars in this context has been studied in \cite{Rinaldi:2012vy1}.

It is important to note that the derivative coupling of the scalar field to the Einstein tensor acts like an effective cosmological  constant
introducing  a new scale in the theory, which on short distances allows us 
 to find hairy black hole solutions with scalar hair just outside the black hole horizon  in asymptotically flat spacetime \cite{Kolyvaris:2013zfa}. On large distances,  the presence of the derivative coupling acts as a friction term in the inflationary period of the cosmological
evolution  \cite{Amendola:1993uh,Sushkov:2009hk,saridakis,germani}. Moreover, it was found that at the end of
inflation in the reheating period, there is a suppression of heavy particle production  as the derivative coupling is increased. This was attributed to the fast decrease of  kinetic
energy of the scalar field due to its  wild oscillations \cite{Koutsoumbas:2013boa}. Also the reheating period in the presence of the derivative coupling was studied in \cite{Dalianis:2016wpu}.

The above discussion indicates that one of the main effects of the kinematic coupling of a scalar field to the Einstein tensor
is that  gravity  influences strongly  the  propagation of the scalar field compared to a scalar field minimally coupled to gravity. As it is well known,
in the application of gauge/gravity duality to simple  condensed matter systems like the holographic superconductor, or to the understanding of thermalization processes in the quark-gluon plasma, it is important to understand in the gravity sector the formation of a black hole and its backreaction to a matter distribution outside its horizon, usually parametrized by a scalar field. Then, it would be instructive to study the effect of  scalar fields
appearing in the Horndeski theory on holography and compare their effect to minimally coupled scalar fields.

One of the first studies of a scalar field coupled to Einstein tensor, a term belonging to the Horndeski action (\ref{horn}) in connection to condensed matter physics using the AdS/CFT correspondence has been carried out in \cite{Kuang:2016edj}. A holographic dual description of a superconductor was presented, with a charged scalar field kinematically coupled to the Einstein tensor in addition to its minimal coupling to gravity. It was found that this coupling simulates the impurity effects on  real materials. This happens because this coupling acts like a friction term in the gravity sector, while its dual holographic counterpart has the same effect as it slows down the carriers of electric charge. There is also  a growing interest in studying thermalization processes using gauge/gravity duality. The motivation for
such a study  is the study of the quark-gluon plasma produced at
Relativistic Hadron Ion Collider which behaves like a strongly interacting liquid.
 The dynamics of such processes in a far-from-equilibrium  plasma cannot
be described with the standard methods of field theory or
hydrodynamics. It has been proposed then, in
\cite{Garfinkle:2011hm,Garfinkle:2011tc}, that   such a rapid
thermalization can be studied via its gravity dual, the
gravitational collapse.

One crucial effect in the out-of-equilibrium  processes is the thermalization time, i.e. the time in field theory
between the injection of energy at $t=0$ and the formation of the
quark-gluon plasma. Then according to the gauge/gravity duality the goal is  to find
the right   gravity methods to describe these processes. One such gravity
process is the gravitational collapse of a scalar field and a
formation of a black hole. Then, the thermalization time is
identified with the time needed for  the formation of an apparent
horizon.

As we discussed above, the presence of a kinematic coupling of the scalar filed to Gravity effects significantly the kinematics of
the scalar field. It would be very  interesting to see if in a collapsing process the collapsing time is affected by the presence of this coupling.
Also it would be interesting to answer the question of whether  the black holes of the Horndeski theory discussed above can be formed  dynamically  via the gravitational collapse of a scalar field
coupled to the Einstein tensor.  In this work we consider a collapsing model
with a scalar field that, apart from the usual kinetic coupling to
gravity, is also kinematically coupled to the Einstein tensor.

To study the gravitational collapse we introduced in the action (\ref{EGBscalaraction1})  a cosmological constant $\Lambda$.
 We  assumed that the scalar field is homogeneous and the collapsing
region follows a FRW metric. Assuming also that the scalar field is not self-interacting, we studied the collapsing process for the scalar field to reach the singularity and we found an analytic solution of the field equations showing that an apparent horizon is formed which covers the singularity. Also we
found that the scalar field is finite on the horizon by construction.
 Calculating the first and second fundamental forms following \cite{Giambo:2005se}, we matched the internal with the external metric and in this way we showed that the apparent horizon is covered by an event horizon and then a black hole is formed.
Finally we reconstructed a potential, based on a general ansatz for the solution.

The work is organized as follows. In Section \ref{genform} we present the
general formulation of the theory giving the field equations
resulting from the action considered. In Section \ref{collapse} we discuss the
gravitational collapsing process of reaching the singularity and the formation of a black hole. In Section \ref{apprh}  we describe
the
formation of apparent horizon. In Section \ref{bouns} we discuss the formation of boundary surface by  matching of internal metric
with an external Schwarzschild-AdS metric resulting to a formation
of an event horizon.  In Section \ref{shell}  we discuss the collapse of a shell of a scalar field. The collapse of the
scalar field in the presence of a self-interacting potential is discussed in   Section \ref{potential}. Our
conclusions are in Section \ref{conc}. In the Appendices we give an exact analytical solution
of the field equations when $\lambda=0,\  \ V=0,$ and an analytic approximation for times near the singularity formation time for $\lambda \neq 0$.

\section{General Formalism}
\label{genform}

To have a general setup of the theory we consider the action (\ref{EGBscalaraction1}) including a
cosmological constant and a potential for the scalar field. We also
assume that the derivative coupling function $G$ is a constant denoted by $\lambda$,
independent from the scalar field. Then the action reads  \be S =
\int d^4x \sqrt{-g}\left\lbrace \frac{R-2\Lambda}{16\pi G} -
\left[\frac{1}{2} g^{\mu\nu} - \frac{1}{2}\lambda
G^{\mu\nu}\right]\pkm\phi\pkn\phi - V(\phi) \right\rbrace~.
\label{naction} \ee Variation of the action above gives the
Einstein equations \be G_{\mu\nu} + \Lambda g_{\mu\nu} = 8\pi G
[T_{\mu\nu} + \lambda \Theta_{\mu\nu}]~, \label{einthet} \ee
Œwhere \be G_{\mu\nu} \equiv R_{\mu\nu} - \frac{1}{2}g_{\mu\nu}R~,
\ee \be T_{\mu\nu} = \pkm\phi\pkn\phi - \frac{1}{2}
g_{\mu\nu}g^{ab} \pa_a\phi \pa_b\phi- g_{\mu\nu} V(\phi)~, \ee
\begin{align*}
\Theta_{\mu\nu} = &\frac{1}{2} \pkm \phi \pkn \phi R - 2 \pa_a
\phi \pa_{(\mu}\phi {R^a_\nu}_)  + \frac{1}{2} G_{\mu\nu}
(\pa\phi)^2 - \nabla^a \phi \nabla^b\phi R_{\mu a \nu b} -
\nabla_\mu \nabla^a \phi \nabla_\nu \nabla_a \phi \nonumber \\
&+\nabla_\mu \nabla_\nu \phi \square \phi - g_{\mu\nu} \big[ -
\frac{1}{2}\nabla^a \nabla^b \phi \nabla_a \nabla_b \phi +
\frac{1}{2}(\square \phi)^2 - \nabla_a \phi \nabla_b \phi R^{ab}
\big]~,
\end{align*}
while the Klein-Gordon equation reads
\be
\square \phi - V_\phi = 0~,
\ee
 where $V_\phi$ denotes the derivative with respect to $\phi$ and
$\square\phi = (-g)^{-1/2}\pkm\big[(-g)^{1/2} [ g^{\mu\nu} +
\lambda G^{\mu\nu}]\pkn\phi\big]$.

To study the gravitational collapse of the scalar field we assume
that it is only time dependent and  that the collapsing region
follows a  metric in comoving coordinates \be ds^2 = -dt^2 +
a^2(t)\left(\frac{dr^2}{1-k r^2} + r^2 d\theta^2 + r^2\sin^2\theta
d\phi^2\right)~.\label{frw} \ee
We define the dimensionless
quantities \bea t && \equiv M_{pl} \cdot t \quad , \quad
\lambda \equiv \lambda \cdot M^2_{pl} \quad , \quad
k \equiv \frac{k}{M^2_{pl}} \quad , \quad
\Lambda \equiv \frac{\Lambda}{M^2_{pl}} \quad ,
\nonumber \\ && \quad \phi \equiv \frac{\phi}{M_{pl}} \quad ,
\quad V(\phi) = \frac{V(\phi)}{ M^4_{pl}} \quad , \quad
V_\phi = \frac{V_\phi(\phi)}{
M^3_{pl}}~. \label{notations}\eea
Then the Klein-Gordon equation
becomes \be \label{eom1} \left(1+ \frac{3\lambda (\dot{a}^2(t) +
k)}{a^2(t)}\right)\ddot{\phi}(t) + \left(3\frac{\dot{a}(t)}{a(t)}
+ 3\lambda\left(\frac{\dot{a}^3(t)}{a^3(t)} +
2\frac{\dot{a}(t)\ddot{a}(t)}{a^2(t)} + \frac{\dot{a}(t) k
}{a^3(t)} \right) \right)\dot{\phi}(t) = - V_\phi~. \ee The
tt-component of the Einstein equations reads \be \label{fr1}
\frac{3(k + \dot{a}^2(t))}{a^2(t)} - \Lambda =
4\pi\left\lbrace \left[1 +
\lambda\left(6\frac{\dot{a}^2(t)}{a^2(t)} + 3\frac{(k+
\dot{a}^2(t))}{a^2(t)} \right)\right]\dot{\phi}^2(t) +
2V(\phi)\right\rbrace~. \ee

In the next section we study in detail the two equations (\ref{eom1}) and (\ref{fr1}).

\section{Gravitational collapse with zero potential}
\label{collapse}

We assume that the collapsing space is flat, $k=0$, and
we first consider the case  in which the scalar field does not
have a self-interaction term, $V(\phi)=0$.  Then equations
\eqref{eom1} and \eqref{fr1} become
 \begin{align} \label{neom} &\left(1+ \frac{3\lambda \dot{a}^2(t)}{a^2(t)}\right)\ddot{\phi}(t) + \left(3\frac{\dot{a}(t)}{a(t)}
+ 3\lambda\left(\frac{\dot{a}^3(t)}{a^3(t)} +
2\frac{\dot{a}(t)\ddot{a}(t)}{a^2(t)} \right) \right)\dot{\phi}(t) =0~. \\[0.1cm]
&\frac{3 \dot{a}^2(t)}{a^2(t)} - \Lambda =
4\pi \left(1 +
9\lambda\frac{\dot{a}^2(t)}{a^2(t)} \right)\dot{\phi}^2(t)~.  \label{EinsteinT}
\end{align}

\begin{figure}[ht!]
\centering \subfloat[] [Scale $a(t)$ for various
$\lambda$. The singularity is reached when $a(t) =
0$.] {
\includegraphics[scale=0.5]{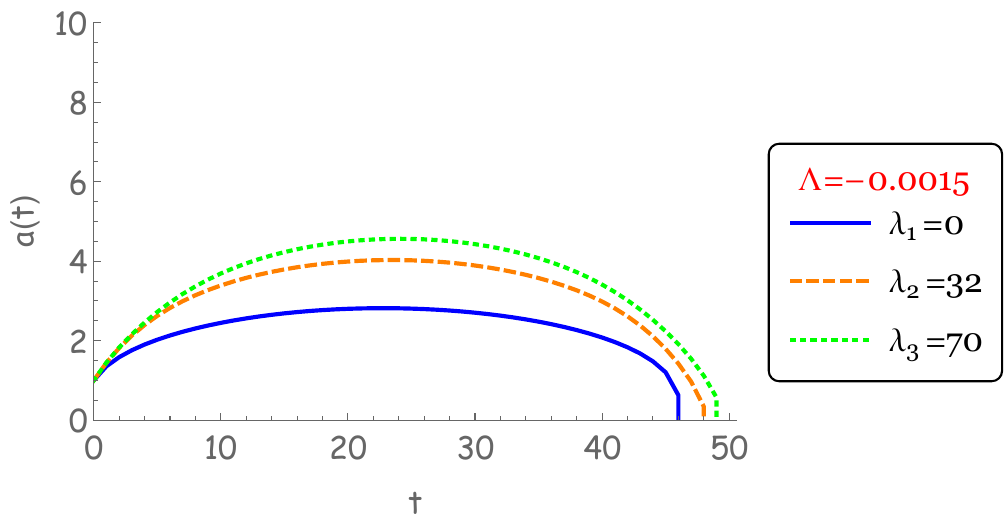}}
\qquad \centering \subfloat[][Derivative $\dot{a}(t)$. Collapse
is starting when $\dot{a}(t) = 0$.]{
\includegraphics[scale=0.5]{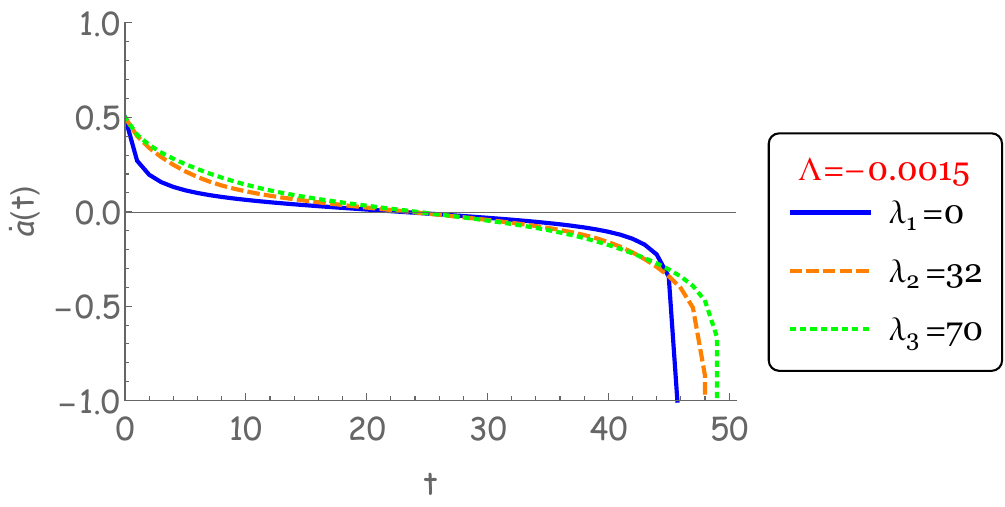}}
\caption{The behaviour of the scale factor.}
 \label{lamda1}
\end{figure}

Note that since the action (\ref{naction}) is invariant under a
shift symmetry (when there is no potential),  only derivatives of
$\phi$ appear in the field equations. We solve equation
\eqref{EinsteinT} with respect to $\dot{\phi}(t)$ and
substitute it into equation \eqref{neom}. The resulting equation
is a function of only $a(t)$ and its derivatives and we solve
this equation numerically. We depict the behaviour of the scale
factor $a(t)$ in Fig.~\ref{lamda1}. The singularity is
reached at the singularity time $t_s$ where $a(t_s) = 0$,
and the collapse starts when $\dot{a}(t) < 0 $. The analytic
solution in Appendix A proves  beyond doubt (for the special case $\lambda=0,\ V=0),$ that the numerical
results, in particular the divergence of $\dot{a}(t)$ at
$t_s$ are not just numerical artifacts. The results in Appendix A are extended in Appendix B to also include the case $\lambda\ne 0,\ \ V=\lambda \dot{\phi}^2.$ To see the behaviour of
the scalar field $\phi$ we substitute the numerical solution for
the scale factor $a(t)$ to \eqref{neom}. The behaviour of the scalar field  appears in a subsequent section.
In our numerical solutions we have chosen the
parameters so that the weak energy condition $\rho(t)
+ 3p(t) \geq 0$, is satisfied, where
\begin{align}
\rho(t) &= T_{00} +\lambda\Theta_{00} = \frac{\dot{\phi}^2(t)}{2} - \frac{9}{2}\lambda\frac{\dot{a}^2(t)}{a^2(t)}~, \label{rho} \\
p(t) &= T_{ii} + \lambda\Theta_{ii} = \frac{\dot{\phi}^2}{2}  - \frac{\lambda\left(\dot{a}^2(t)\dot{\phi}^2(t) + 2a(t)\dot{\phi}(t)\left(\dot{\phi}(t)\ddot{a}(t) + 2\dot{a}(t) \ddot{\phi}(t)\right)\right)}{2a^2(t)}. \label{pres}
\end{align}\\
\begin{figure}[ht!]
\begin{center}
\includegraphics[scale=0.5]{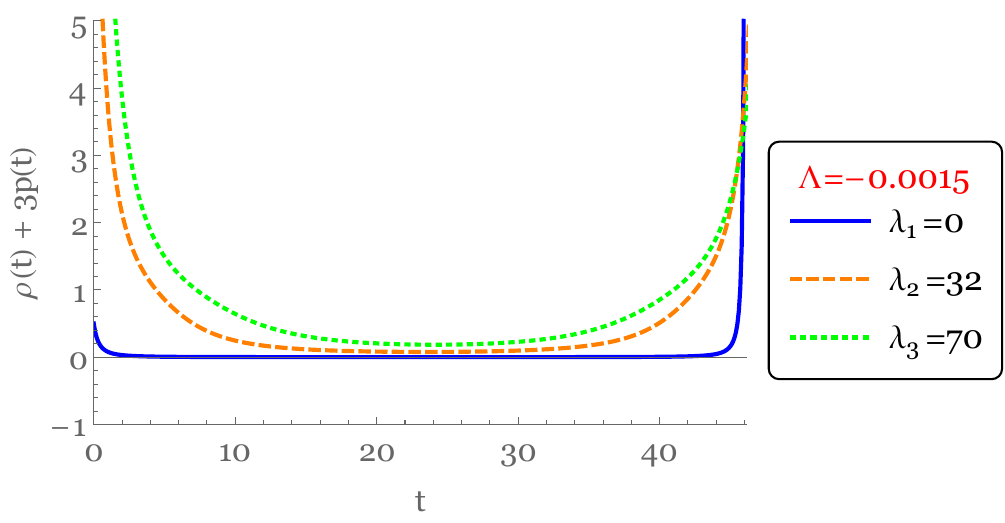}
\end{center}
\caption{Weak energy condition  $\rho(t) + 3 p(t) \geq 0$ satisfied for various $\lambda$.}
\label{cond1}
\end{figure}
In Fig.~\ref{cond1} we depict the weak energy condition for various values of $\lambda$. As we can see, the weak
energy condition $\rho(t)
+ 3p(t) \geq 0$ is satisfied for all times.

We observe that, as the value of the derivative coupling
$\lambda$ increases, the collapse time as well as the
singularity formation time are also increasing. This behaviour of
the collapsing scale factor can be understood on the basis of the
fact that, in a time dependent background, the derivative coupling
 $\lambda$ acts as a friction
term
\cite{Amendola:1993uh,Sushkov:2009hk,germani,Koutsoumbas:2013boa,saridakis},
so it takes more time to reach the singularity.

\begin{figure}[h!]
\begin{center}
\includegraphics[scale=0.55]{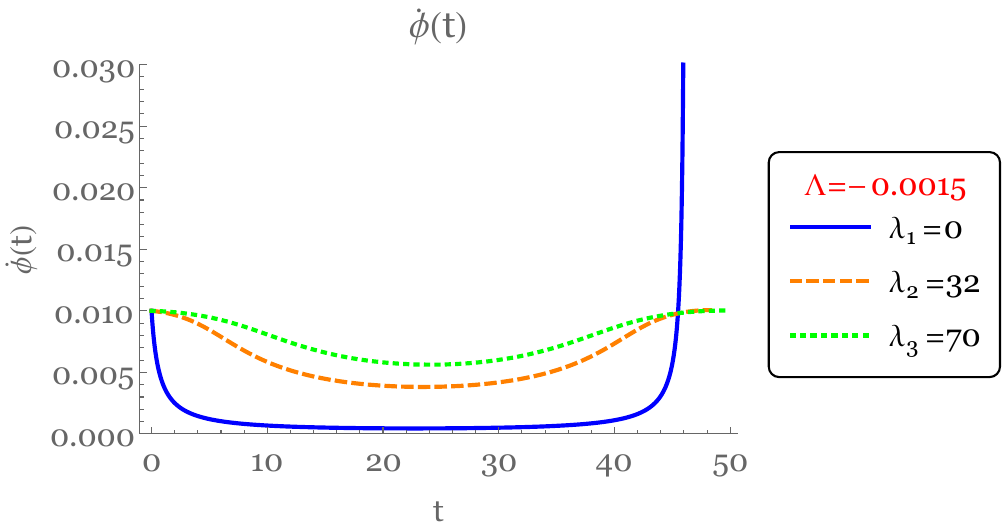}
\end{center}
\caption{Numerical solution of the scalar field derivative, $\dot{\phi}(t)$ for various $\lambda$.}
\label{scalarplot}
\end{figure}

Another interesting effect is that at $t = t_{\text{singularity}}$ for $\lambda = 0$ the scalar field derivative blows up while in the presence of the non-minimal coupling it takes finite values. We can see that by solving \eqref{neom},\eqref{EinsteinT} with respect to $\dot{\phi}(t)$ using the numerical solution of $a(t)$. This is shown in Fig. \ref{scalarplot}
\\
\subsection{Formation of apparent horizon}
\label{apprh}.

To study the formation of a naked singularity or  a black hole
 we need to investigate the dynamics of apparent
horizons. We analyze under which conditions trapped surfaces
form and investigate their properties in connection with the
derivative coupling $\lambda$.

\begin{figure}[ht!]
\centering \subfloat[][Scale $a(t)$ for various
$\lambda$ without the derivative coupling
$(\lambda = 0)$]{
\includegraphics[scale=0.45]{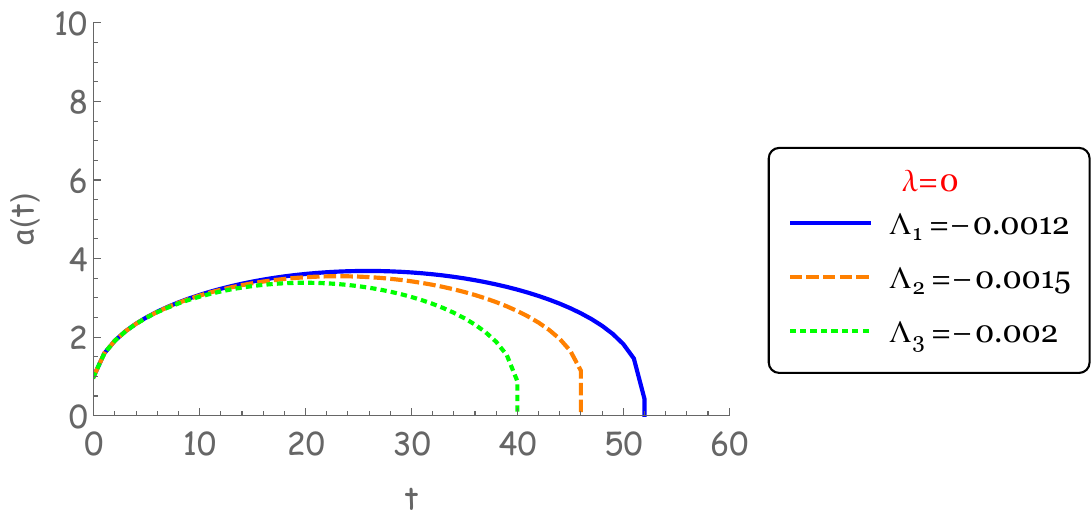}}
\qquad
\centering
\subfloat[][$\dot{a}(t)$ for various $\Lambda$. $(\lambda =  0)$]{
\includegraphics[scale=0.45]{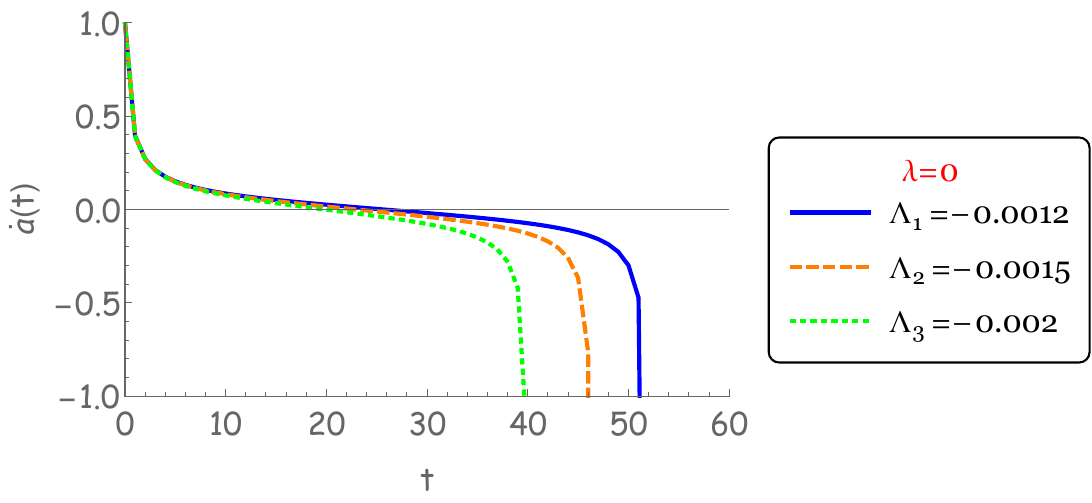}}
\caption{Formation of the singularity in the presence of a
cosmological constant, with no derivative coupling $(\Lambda =  0).$} \label{fig111}
\end{figure}

 Let us first discuss the case of the derivative coupling of the scalar field with $\lambda=0$. In this case
 the energy momentum tensor becomes
\beno
T_\mu^\nu = \text{diag}(-\epsilon,p,p,p)~,
\eeno
with
\be \label{denpre}
 \epsilon = \frac{1}{2}\dot{\phi}^2 + V(\phi) \quad \text{and} \quad p = \frac{1}{2}\dot{\phi}^2 - V(\phi)~.
\ee
Working with dimensionless quantities the field equations reduce to the familiar forms
\begin{align}
&\text{Friedmann}: \qquad \dot{a}^2(\tau) = \frac{8\pi}{3}a^2(t) \epsilon \label{fr}~, \\
&\text{Klein-Gordon}: \qquad \ddot{\phi}(t) + 3\frac{\dot{a}(t)}{a(t)}\dot{\phi}(t)= -V_\phi(\phi)~.
\end{align}
For $V(\phi) = 0 $ the above equations can be solved numerically for different values of $\Lambda$ as we show in Fig. \ref{fig111}.\\
\\
In Appendix A we find an analytic solution for this case
\be
a(t) = \left(\frac{\cosh\left(\sqrt{3
\Lambda} (t - C_1)\right)}{\cosh\left(\sqrt{3
\Lambda} C_1\right)}\right)^{1/3}~,
\ee
We mention here that for positive cosmological constant the scale $a(t)$ does not vanish so the field does not collapse to a singularity. For $\Lambda < 0$ though we get
\be \label{ana}
a(t) = \left(\frac{\cos\left(\sqrt{3
|\Lambda|} (t - C_1)\right)}{\cos\left(\sqrt{3
|\Lambda|} C_1\right)}\right)^{1/3}~.
\ee
Then it is evident that
\beno a\left(t
=C_1+\frac{\pi}{2 \sqrt{3 |\Lambda|}}\right)=0~,
\eeno


so the singularity forms at a  finite  time $t_s =C_1+\frac{\pi}{2 \sqrt{3 |\Lambda|}}$.
By differentiating \eqref{ana} we find
$$\dot{a}\left(t =C_1+\frac{\pi}{2 \sqrt{3 |\Lambda|}}\right) =\frac{\sqrt{|\Lambda|}}{\sqrt{3}\cos\left(\sqrt{3\Lambda}(C_1)\right)}
\left. \frac{\sin\left(\sqrt{3 |\Lambda|} (t-C_1)\right) }
{\cos^{2/3}\left(\sqrt{3|\Lambda|}
(t-C_1)\right)}\right|_{t =C_1+\frac{\pi}{2 \sqrt{3
|\Lambda|}}} = -\infty~.$$
We want to investigate whether the singularity is covered by an apparent horizon. If we denote by $t_h(r)$ the apparent horizon curve, then the boundary
of this curve is defined by the relation $R(r,t_h(r)) = 2m(r,t_h(r))$, where $m(r,t_h(r))$ is the Misner-Sharp mass and $R(r,t) \equiv r a(t).$  Then, any surface with coordinates
$ (r,t)$ inside the trapped region satisfies the relation \beno
\mathcal{T}= \lbrace (r,t) : R(r,t) \leq 2m(r,t) \rbrace ~.
\eeno
For a FRW metric the inequality above gives
\beno
2m(r,t) = 2R\left(1 - g^{\alpha\beta}\pa_\alpha R\pa_\beta R\right) =r^3 \dot{a}^2(t)a(t)~.
\eeno
Therefore for any surface to belong to the trapped region the condition $r^2 \dot{a}^2 \geq 1 $ must hold. We have found that
$\dot{a}^2(t_s) = + \infty,$ therefore a time $t_h (r) < t_s$ there must exist, such that $\dot{a}^2$ takes its minimum value  $\dot{a}^2(t_h(r)) = \frac{1}{r^2}.$
Then the apparent horizon lies below the singularity curve $t=t_s$, which is therefore covered for any $r > 0$.  So for negative cosmological constant and zero potential we have the formation of an apparent horizon. In agreement with \cite{Giambo:2005se}, \cite{Baier:2014ita} the shell of the scalar field becomes trapped before it becomes singular, and so a black hole forms. On the other hand, as we will show in Section \ref{potential}, if $\dot{a}(t_s)$ is finite we are able to find a $r = r_b > 0$ such that no apparent horizon forms.

 Let us now switch on the   derivative coupling of the scalar field  $\lambda \neq 0$.
 The field equations now become
\be \label{frW}
3\frac{\dot{a}^2}{a^2}\left(1-12\pi \lambda \dot{\phi}^2\right) - (\Lambda + 8\pi V(\phi) + 4\pi \dot{\phi}^2) = 0~,
\ee
\be
\frac{3\dot{a}(t)\dot{\phi}(t)+a(t)\ddot{\phi}(t)}{a(t)} + \frac{3\lambda\dot{a}\left(\dot{a}^2(t)\dot{\phi}(t) + 2a(t) \dot{\phi}^2(t)\ddot{a}(t)+a(t)\dot{a}(t) \ddot{\phi}(t)\right)}{a^3(t)} = - V_\phi(\phi)~.
\ee

\begin{figure}[h!]
\centering
\includegraphics[scale=0.5]{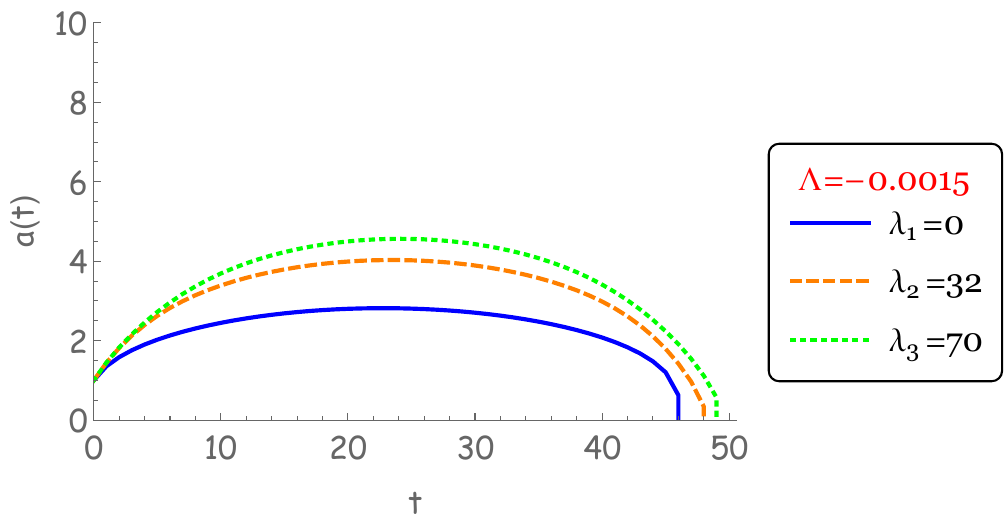}
\caption{numerical  solution of $a(t)$ for $\lambda > 0, \Lambda  = -0.0015$}  \label{fig333}
\end{figure}

For $V(\phi) = 0$ the above equations can be solved numerically. Fixing the cosmological constant $\Lambda = -0.0015$ we get the function $a(t)$, which is displayed in Fig. \ref{fig333}. \\
\\
It is important to show that $\dot{a}(t)$
actually goes to $-\infty$ at $t = t_s$ and check whether this property depends on $\lambda.$ In Appendix B we solve the
equations analytically at times close to the singularity time with the results
\be
a(t) \sim (t_s - t)^{\frac{2}{3}} \quad \Rightarrow \quad \dot{a}(t) \sim -(t_s - t)^{-\frac{1}{3}}
\ee
Thus the scale factor
vanishes for a suitable value of $t = t_{s}$ and its
derivative diverges to $-\infty$, since the exponent $-\frac{1}{3}$ is negative, signaling the formation of an apparent
horizon and thus a black hole. The above results are
$\lambda$ independent, so the apparent horizon
is formed for any value of the derivative coupling.

Another interesting effect which can be seen in Fig. \ref{fig111}
is that as the absolute value of the cosmological constant
increases, the collapse and singularity formation time
decrease. This is an opposite effect of what we had observed in
the case of the increase  of the value of the derivative
coupling \cite{Koutsoumbas:2013boa}.
The cosmological constant $\Lambda = -3/l^2$ defines a scale  $l$
in the $AdS_4$ space, therefore the decrease of the collapse and
singularity time may be understood as the result of the scalar
field having less distance to travel through the $AdS_4$ space.


Summarizing, we found that the apparent horizon is formed for any
value of the derivative coupling $\lambda$ protecting
the singularity. The derivative coupling $\lambda$ enhances the action of the cosmological
constant, indicating that the derivative coupling
$\lambda$ acts as an effective cosmological constant. This has been noticed first in \cite{Rinaldi:2012vy}. Other relevant publications include
\cite{Amendola:1993uh,Sushkov:2009hk,germani,Koutsoumbas:2013boa,saridakis}.


\subsection{Formation of a boundary surface}
\label{bouns}

To construct the appropriate event horizon that signals the
formation of a  black hole we match the FRW metric of our collapsing
scalar field
\be \label{flatfrw}
ds_{int}^2 = -dt^2 + a^2(t)dr^2 + r^2 a^2(t)d\Omega~,
\ee
with that of a Schwarzschild-$AdS_4$ metric
\be \label{schwads4}
ds_{ext}^2 = - \chi(Y) dT^2 + \chi(Y)^{-1} dY^2 + Y^2 d\Omega^2,\qquad \chi(Y) = 1 - \frac{2 M(Y)}{Y} + \frac{Y^2}{l^2}~,
\ee
at a hypersurface $\Sigma$. We discuss the formalism following mainly \cite{Giambo:2005se} and then we apply it to our
study. We parametrize $\Sigma$ with coordinates $y^a = (\tau,\theta,\phi)$. The injection of $\Sigma$ into the {\it internal} space equipped with metric \eqref{flatfrw} reads  $y^a =(\tau,\theta,\phi) \hookrightarrow x^{\alpha} =(\tau,r_b,\theta,\phi)$. The first fundamental form $($induced metric$)$ is defined by
\be
I^{int}_\Sigma = \underbrace{g_{\alpha\beta} \frac{dx^\alpha}{dy^a}\frac{dx^\beta}{dy^b} }_{h_{ab}} dy^a dy^b~,\label{firstff}
\ee
from which we get
\be \label{indint}
I^{int}_\Sigma = -d\tau^2 + r_b^2 a^2(t) d\Omega^2~.
\ee
The normal vector of $\Sigma$ according to \eqref{flatfrw} satisfies the relation $g_{\alpha\beta}\frac{\partial x^\alpha}{\partial y^a}\xi^\beta = 0$
from which we get
  $\xi^\alpha = (0,\xi^1,0,0)$ and $\xi_{\alpha} = (0,\xi^1 a^2(t),0,0)$. The normalized normal vector with $n^\alpha n_\alpha = 1$ $($spacelike vector$)$ is
\beno
n^\alpha = (0,a^{-1}(t),0,0)~.
\eeno
The Second fundamental form  $($the extrinsic curvature$)$ of the hypersurface $\Sigma$ is given by
\be \label{extcurv}
K_{ab} = - n_\alpha\left(\frac{\pa^2 x^\alpha}{\pa y^a\pa y^b} + \Gamma^\alpha_{\rho\sigma} \frac{\pa x^\rho}{\pa y^a} \frac{\pa x^\sigma}{\pa y^b}\right)~,
\ee
which for the normal vector $n^\alpha$ of \eqref{flatfrw} gives
\be \label{extint}
II^{int}_\Sigma = 0\cdot d\tau^2 + r_b a(t) d\Omega^2~.
\ee
Performing the same operation with the {\it external} metric \eqref{schwads4} the injection reads in coordinates $y^a =(\tau,\theta,\phi) \hookrightarrow x^{\alpha} =(T(\tau),Y(\tau),\theta,\phi)$. Then from the  first fundamental form (\ref{firstff}) we find
\be \label{indext}
I^{ext}_\Sigma = \left(-\chi(Y) \dot{T}^2+ \frac{1}{\chi(Y)}  \dot{Y}^2\right) d\tau^2 +   Y^2(\tau) d\Omega^2~.
\ee
Matching the first fundamental forms \eqref{indint} and \eqref{indext} we get
\be \label{match1}
\left(-\chi(Y) \dot{T}^2 + \frac{\dot{Y}^2}{\chi(Y)}\right) =-1
\ee
and
\be \label{match2}
Y^2(\tau) = r_b^2 a^2(t)~.
\ee
The normal vector of \eqref{schwads4} is found to be
\beno
\xi^\alpha = \left(\xi^0,  - \frac{g_{00}\dot{T}}{g_{11}\dot{Y}}\xi^0,0,0\right)~, \eeno
while the normalized normal vector is
\beno
n^\alpha = \left(\frac{1}{\sqrt{-\chi(Y) + \frac{\chi^3(Y)\dot{T}^2}{\dot{Y}^2}}},\frac{\chi^2(Y)\dot{T}(Y)}{\sqrt{-\chi(Y) + \frac{\chi^3(Y)\dot{T}^2}{\dot{Y}^2}}\dot{Y}},0,0\right)
\eeno
from which using (\ref{extcurv}) we get the second fundamental
\be \label{extext}
II^{ext}_\Sigma = -\frac{\chi^2(Y)\chi(Y)_{,Y}\dot{T}^3 + 3   \chi(Y)_{,Y}\dot{T}\dot{Y}^2 + 2\chi(Y) \left( \dot{Y}\ddot{T} - \dot{T}\ddot{Y}\right)}{2\chi(Y)}d\tau^2 +  Y \chi(Y) \dot{T} d\Omega^2~.
\ee
Matching the second fundamental forms of the internal space \eqref{extint} and the external space \eqref{extext} we get
\be \label{match3}
-\frac{\chi^2(Y)\chi(Y)_{,Y}\dot{T}^3 + 3   \chi(Y)_{,Y}\dot{T}\dot{Y}^2 + 2\chi(Y) \left( \dot{Y}\ddot{T} - \dot{T}\ddot{Y}\right)}{2\chi(Y)} = 0~,
\ee
and
\be \label{match4}
  Y \chi(Y) \dot{T}  = r_b a(\tau)~.
\ee
Equations \eqref{match1}, \eqref{match2}, \eqref{match3} and \eqref{match4} completely specify the matching conditions at the boundary of the collapsing scalar field. We can check that \eqref{match3} holds identically by using the remaining three conditions $($\eqref{match1} , \eqref{match2} and \eqref{match4}$)$. In addition \eqref{match1} says that the Misner-Sharp mass is continuous across $\Sigma$, that is $2m(r_b,\tau) = Y_b(\tau)\left(1-\chi(Y_b)\right)$. From that we get a time-dependent mass
\be \label{tmass}
r_b^3 \dot{a}^2(\tau)a(\tau) = r_b a(\tau)\left(1-\left(1-\frac{2M(\tau)}{r_b a(\tau)} + \frac{r_b^2 a^2(\tau)}{l^2}\right)\right) \quad \Rightarrow \quad M(\tau) =\frac{1}{2}r_b^3 a(\tau)\left(\dot{a}^2(\tau) + \frac{a^2(\tau)}{l^2}\right)~.
\ee
From \eqref{match4} we see that the time coordinate must satisfy
\beno
\frac{dT}{d\tau} = \frac{1}{\chi(Y_b)}~,
\eeno
so the singularity formation time in the outside coordinate is given by
\be \label{sintime}
T_s = \int_0 ^{\tau_s} \frac{d\tau}{\chi(Y_b)} \stackrel{\eqref{tmass}}{=} \int_0 ^{\tau_s} \frac{d\tau}{1-r_b^2 \left(\dot{a}^2 + \frac{a^2}{l^2}\right) + \frac{r_b^2 a^2}{l^2}} = \int_0 ^{\tau_s} \frac{d\tau}{1-r_b^2 \dot{a}^2}~.
\ee
We see that because $\dot{a}^2(\tau)$ is not bounded in $(0,\tau_s)$ we cannot find a $r_b > 0$ so that $1 - r_b^2\dot{a}^2$ is bounded away from zero. Thus the function $\frac{1}{\chi(Y_b)}$ is not integrable in $(0,\tau_s)$. For radial null geodesics we set $ds^2=0$ on the $AdS_4$-Schwarzschild metric and we get
\beno
0 = -\chi(Y)dT^2 + \frac{dY^2}{\chi(Y)} \quad \Rightarrow \quad \frac{dT}{dY} = \frac{1}{\chi(Y)}~.
\eeno
But as we argued the above ordinary differential equation has no solution, so a radial null geodesic starting from the singularity does not exist. The boundary $\Sigma = \lbrace r = r_b \rbrace$ of the scalar field collapses to a black hole.

\subsection{Behaviour of the scalar field shell}
\label{shell}

\begin{figure}[h!]
\centering
\includegraphics[scale=0.50]{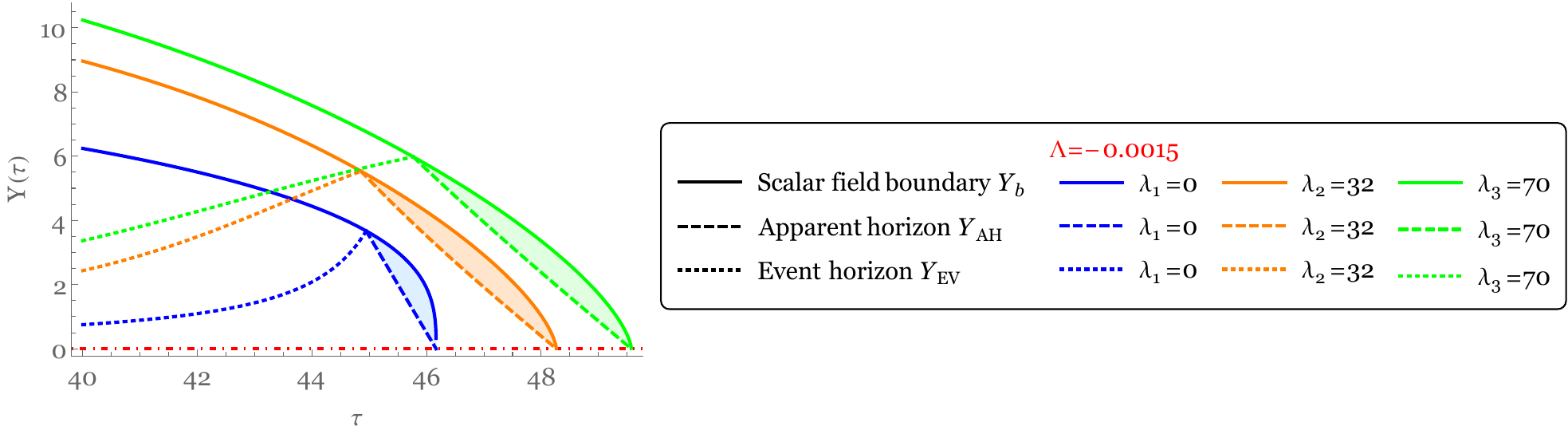}
\caption{The event horizon $($dotted$)$ grows until it coincides with the boundary of the scalar field sphere. At this time the apparent horizon $($dashed$)$ forms  and shrinks to zero size as the boundary of the collapsing shell reaches the singularity. The trapped region lies between the boundary and the apparent horizon.} \label{boundary}
\end{figure}

\begin{figure}[h!]
\centering \subfloat[][Trapped Area versus $\lambda$.]{
\includegraphics[scale=0.50]{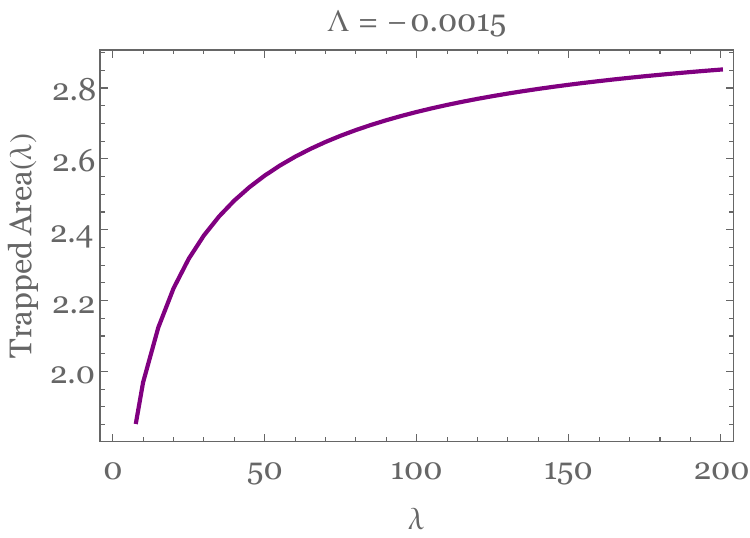}  \label{trapped}} \qquad \qquad
\centering \subfloat[][Trapped boundary mass for $\lambda = 6\; ($thick$) $ , $\lambda = 11 \;($dashed$) $ , $\lambda = 26\; ($dashed-dooted$)$]{
\includegraphics[scale=0.50]{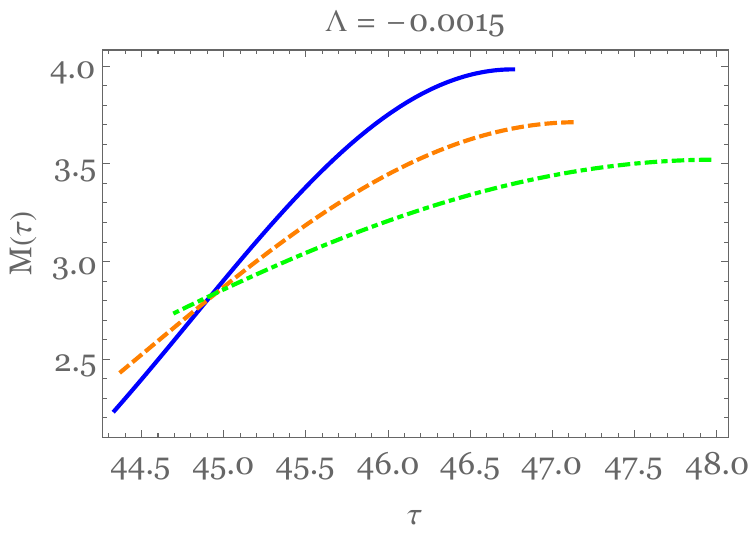}\label{formation}}
\caption{Trapped area and boundary mass for various values of the derivative coupling $\lambda$.} \label{massf}
\end{figure}

\begin{figure}[h!]
\begin{center}
\includegraphics[scale=0.4]{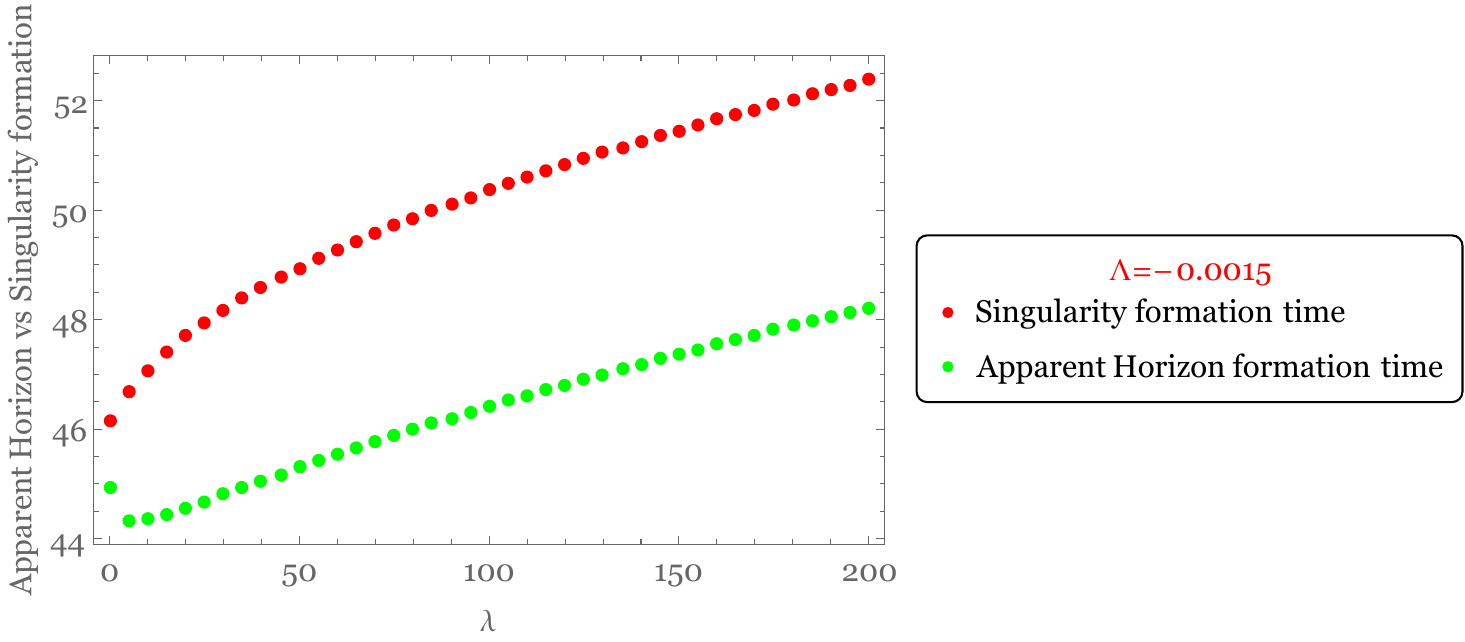}
\end{center}
\caption{Formation time for the apparent horizon and singularity versus $\lambda$.} \label{vslambda}
\end{figure}

In order to study the behaviour of the scalar field shell we have to express the metrics \eqref{flatfrw}, \eqref{schwads4} in the same coordinates.   Following \cite{Baier:2014ita}, we do this by choosing the radial coordinate for the interior spacetime \eqref{flatfrw} to be
\beno
Y = r a(t) \quad \Rightarrow \quad a(t) dr = dY - r \dot{a}(t) dt
\eeno
leading to
\be \label{frwY}
ds^2 = \left(\frac{Y^2\dot{a}^2(t)}{a^2(t)} -1 \right) dt^2 + dY^2 - 2\frac{\dot{a}(t)}{a(t)}Y dt dY + Y^2d\Omega^2~.
\ee
For the exterior region we use the Painleve - Gullstrand transformation
\beno
T = t + f(Y)
\eeno
leading to
\beno
ds^2 = -\chi(Y) dt^2 - \left(\chi(Y) f'^{{}^2}(Y) - \chi^{-1}(Y)\right)dY^2 - 2\chi(Y) f'(Y) dtdY + Y^2 d\Omega^2~,
\eeno
where $f'(Y) = \frac{df(Y)}{dY}$. We can choose $f'(Y)$ so that the term multiplying $dY^2$ is unity. That is $ f'^{{}^2}(Y) = \frac{\chi^{-1}(Y) - 1}{\chi(Y)}$. Then the metric becomes
\be \label{schwads4Y}
ds^2 = -\chi(Y) dt^2 + dY^2 - 2\chi(Y) \sqrt{ \frac{\chi^{-1}(Y) - 1}{\chi(Y)}}dt dY + Y^2 d\Omega^2~.
\ee
Matching the temporal parts of the metrics \eqref{frwY} and \eqref{schwads4Y} we get the matching condition
\beno
 \left(\frac{Y^2\dot{a}^2(t)}{a^2(t)} -1 \right) =  -\chi(Y) \quad \Rightarrow \quad \frac{\dot{a}^2(t)}{a^2(t)} = \frac{2M(t)}{Y^3}  -\frac{1}{l^2}~.
\eeno
Using the Misner-Sharp expression (\ref{tmass}) for the boundary we get
\beno
Y_b(\tau) = r_b a(\tau)~,
\eeno
where $Y_b$ denotes the boundary of the collapsing scalar field shell. We can also find the {\it apparent horizon} behavior by setting the temporal term $  \left(\frac{Y_{AH}^2\dot{a}^2(t)}{a^2(t)} -1 \right)  = 0$. Finally we can find the {\it event horizon} from the relation $\left(1 - \frac{2M(t)}{Y_{EV}} + \frac{Y_{EV}^2}{l^2}\right) =0$. As we have already mentioned in section \ref{apprh} since $\dot{a}^2(t_s) = +\infty$ we can choose any $r_b$ and still get a trapped region. In our simulations we use $r_b = 3$. Putting all these together in Fig. \ref{boundary} we show that the apparent horizon forms and shrinks to zero size as the boundary of the collapsing shell reaches the singularity. The trapped region lies between the boundary and the apparent horizon. Its appearance changes, but some related quantities depend monotonically on the non-minimal coupling $\lambda.$ In Fig. \ref{trapped} we depict the area of the trapped region versus $\lambda,$ while in Fig. \ref{formation} one may see the trapped boundary mass as a function of time for various values of $\lambda.$ The formation time of the apparent horizon and the singularity versus  $\lambda$ is shown in Fig. \ref{vslambda}. In Fig. \ref{formation} we show that the apparent horizon is always formed before the singularity. The shell of the scalar field becomes trapped before it becomes singular, and so a black hole forms.


\section{Gravitational collapse with non-zero potential}
\label{potential}

Including a potential term in our theory
\be S =
\int d^4x \sqrt{-g}\left\lbrace \frac{R-2\Lambda}{16\pi G} -
\frac{1}{2} g^{\mu\nu}\pkm\phi\pkn\phi + \frac{1}{2}\lambda
G^{\mu\nu}\pkm\phi\pkn\phi - V(\phi) \right\rbrace~,
\ee
we get the Friedmann equation
\beno
\Lambda + 8\pi V(\phi(t)) + 4\pi \dot{\phi}^2(t) - 3\frac{\dot{a}^2(t)}{a^2(t)}\left(1-12\pi \lambda \dot{\phi}^2(t)\right) = 0~,
\eeno
and the Klein - Gordon equation
\beno
-V_\phi(\phi(t))  -\dot{\phi}(t) \left(3\frac{\dot{a}(t)}{a(t)} + 3\lambda \frac{\dot{a}^3(t)}{a^3(t)} + 6\lambda \frac{\dot{a}(t)\ddot{a}(t)}{a^2(t)}\right) - \left(1 + 3\lambda\frac{\dot{a}^2(t)}{a^2(t)}\right) \ddot{\phi}(t) = 0~.
\eeno
The energy density and the pressure of the  energy-momentum tensor
\beno
T_\mu^\nu = \text{diag}(-\epsilon,p,p,p)~,
\eeno
are
\be \label{deden}
 \epsilon = \frac{\dot{\phi}^2}{2} + V(\phi) + \frac{9}{2}\lambda\frac{\dot{a}^2(t)}{a^2(t)}\dot{\phi}^2(t)~,
 \ee
 and
\be \label{depre}
p = \frac{\dot{\phi}^2}{2} - V(\phi) - \frac{\lambda\left(\dot{a}^2(t)\dot{\phi}^2(t) + 2a(t)\dot{\phi}(t)\left(\dot{\phi}(t)\ddot{a}(t) + 2\dot{a}(t) \ddot{\phi}(t)\right)\right)}{2a^2(t)}~.
\ee
We can write the Friedmann equation in the form
\be \label{frA}
\frac{\dot{a}^2(t)}{a^2(t)} - \frac{\Lambda}{3} = \frac{8\pi}{3}\epsilon~.
\ee
Then, following the method of \cite{Giambo:2005se} and \cite{Joshi}, in order to study a collapsing scenario we demand the energy density to follow the relation
\beno
\epsilon = a^{-\nu}(t)~, \quad \nu > 0~.
\eeno
Substituting the above relation to  \eqref{frA} we get
\beno
\dot{a}^2(t) = \frac{\Lambda}{3}a^2(t) + \frac{8\pi}{3}a^{2-\nu}(t)~,
\eeno
or
\beno
\dot{a}^2(t) = \frac{\Lambda}{3}a^2(t) + \frac{8\pi}{3}a^{2\beta}(t)~,\qquad \text{with} \; \beta = 1-\frac{\nu}{2}~,
\eeno
where we have introduced the parameter $\beta$ as in \cite{Baier:2014ita}. Using the relation  $\frac{d}{dt}\left(\epsilon a^3\right) = - p \frac{d}{dt}\left(a^3\right)$ for $\epsilon = a^{-\nu}$ we find that $p = - \frac{3-\nu}{3a^\nu}$. Introducing the equation of state parameter $w$ from  $w = \frac{p}{\epsilon}$, and taking in account the bounds of $-1 < w < 1$, we find $0 < \nu < 6$ or $-2 < \beta < 1$. So one has to focus on this range of $\beta$-values.

\begin{figure}[h!]
\centering \subfloat[][ $a(t)$ for $\Lambda=-0.0015$ and various values of  $\beta$.]{
\includegraphics[scale=0.47]{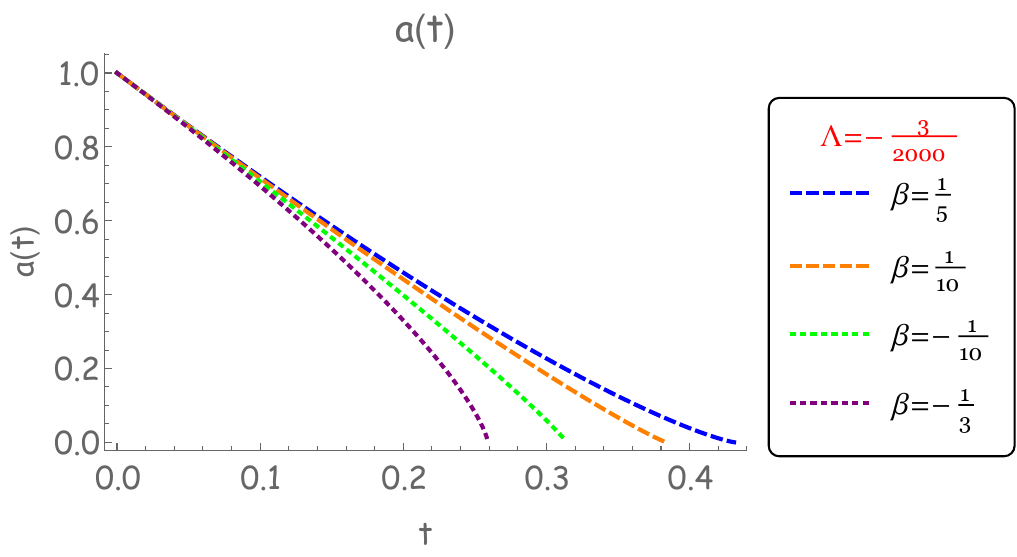}  \label{a}} \qquad \qquad
\centering \subfloat[][$\dot{a}(t)$ for $\Lambda=-0.0015$ and various values of  $\beta$.]{
\includegraphics[scale=0.47]{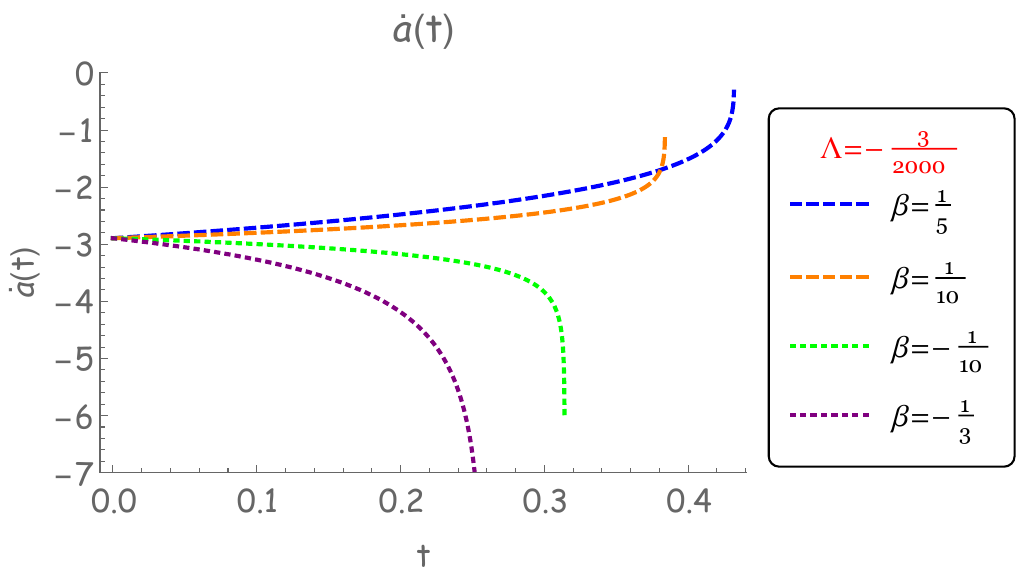}\label{ad}}
\caption{Scale $a(t)$ and its derivative $\dot{a}(t)$ for $\Lambda=-0.0015$ and various values of $\beta$.} \label{aad}
\end{figure}

We take initial data at an initial time, say $t = 0$, to be $a(0) =1$. The field evolves until it reaches a singular state at $a(t_\text{singularity}) = 0$. We consequently consider cases where $a(t)$ is a monotonically decreasing function of $t$. So we solve the equation
\beno
\dot{a}(t) = - \sqrt{\frac{\Lambda}{3}a^2(t) + \frac{8\pi}{3}a^{2\beta}(t)}
\eeno
and we find
\be \label{sola}
a(t) = (8\pi)^{\frac{1}{2-2\beta}} \left\lbrace\sqrt{\frac{1}{|\Lambda|}} \sin\left[\frac{t(\beta-1)}{\sqrt{3}\sqrt{\frac{1}{|\Lambda|}}} - \arcsin \left(\left((8\pi)^{-\frac{1}{2}}\right) \sqrt{\frac{1}{|\Lambda|}}\Lambda \right)\right]\right\rbrace^{\frac{1}{1-\beta}}~.
\ee
The  black hole or naked singularity formation is decided by $\dot{a}(t)$ and then the matching conditions of the fundamental forms have to be checked. This study was carried out in   \cite{Baier:2014ita} for $\Lambda <0$ and it was found found that a naked singularity is formed for $0 < \beta < 1$ and an apparent horizon exists $($black hole formation$)$ for $-2 < \beta < 0$. This setup and results are not affected by the non-minimal derivative coupling. In Fig.\ref{aad} we show the form of the scale $a(t)$ and its derivative $\dot{a}(t)$ for different values of $\beta$. From both solution \eqref{sola} and Fig.\ref{aad}  it's easy to see that $\dot{a}(t)$ is finite for $0 < \beta < 1 $ $($ naked singularity case $)$ while it goes to $-\infty$ for $-2 < \beta < 0$ $($ black hole formation $)$.\\
\\

\begin{figure}[h!]
\centering \subfloat[][ $\phi_{NMC}(t)$ for different values of  $\beta$.]{
\includegraphics[scale=0.47]{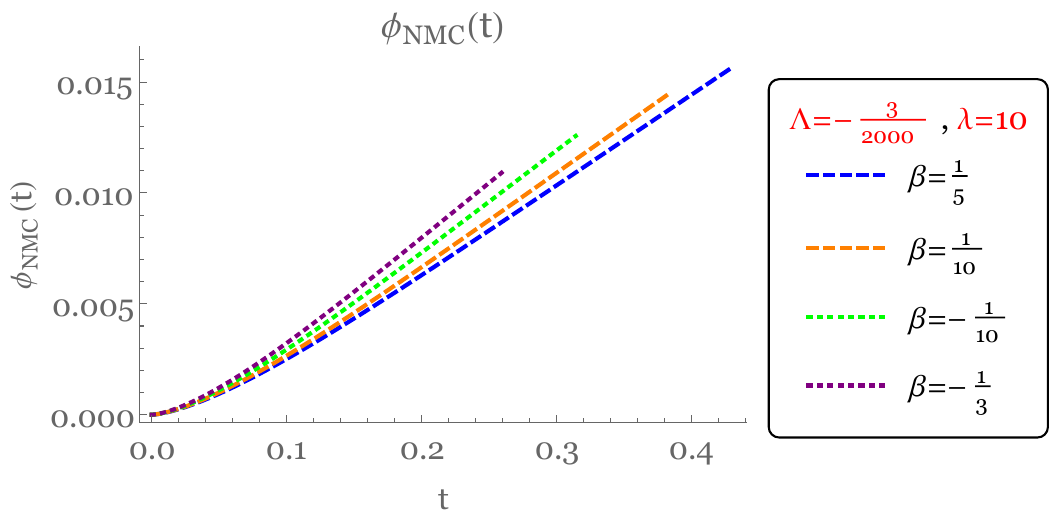}  \label{phiNMC}} \quad \quad
\centering \subfloat[][$\phi_{MC}(t)$ for different values of  $\beta$.]{
\includegraphics[scale=0.47]{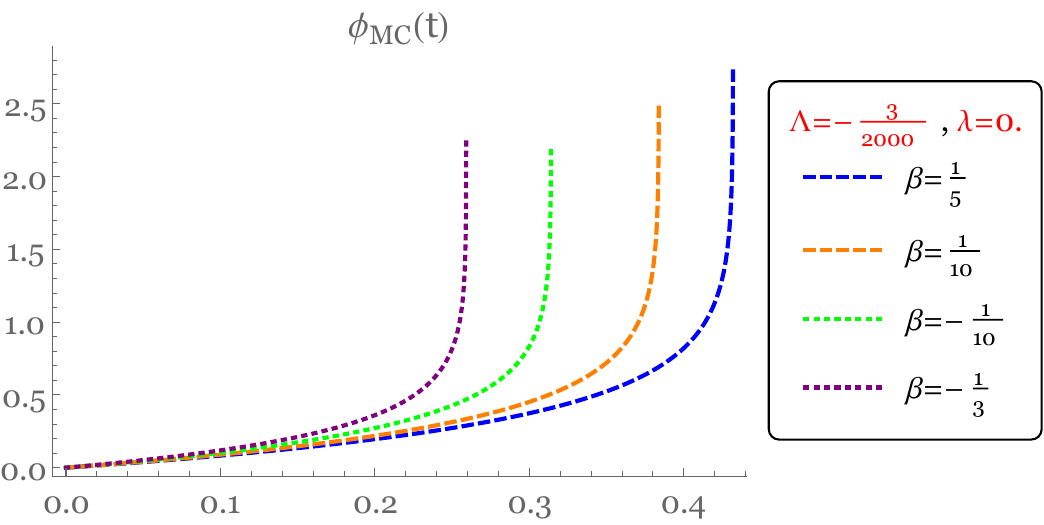}\label{phiMC}}
\caption{Scalar field evolution for different values of $\beta$.} \label{phil}
\end{figure}

\begin{figure}[h!]
\centering \subfloat[][ $\dot{\phi}^2_{NMC}(t)$ for different values of  $\beta$.]{
\includegraphics[scale=0.47]{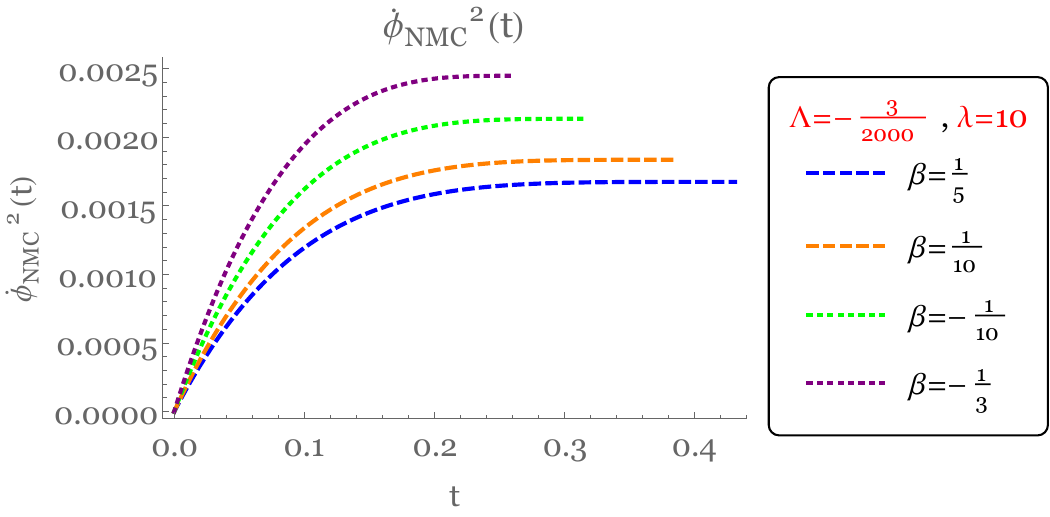}  \label{phidNMC}} \quad \quad
\centering \subfloat[][$\dot{\phi}^2_{MC}(t)$ for different values of  $\beta$.]{
\includegraphics[scale=0.47]{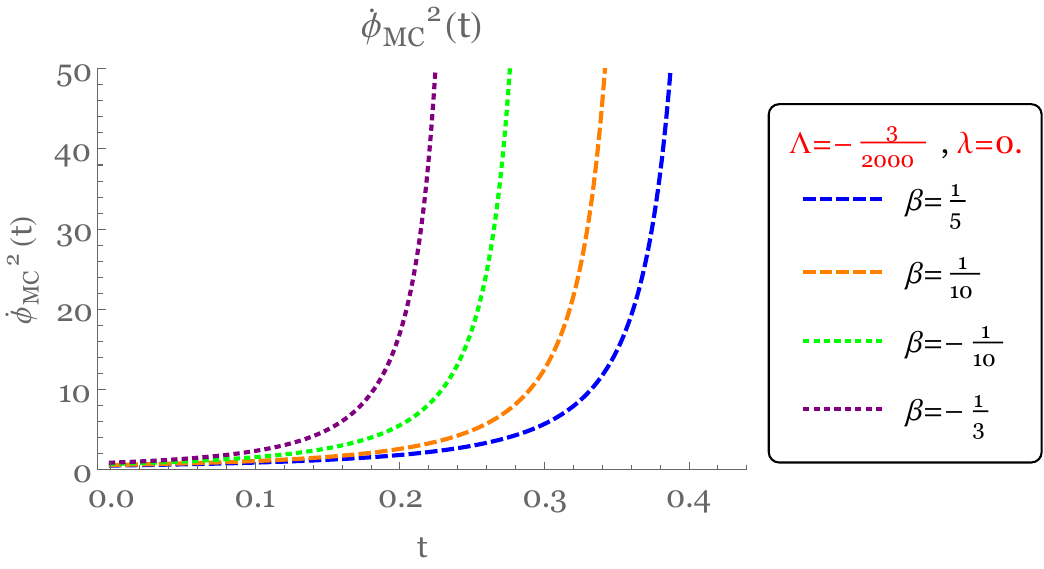}\label{phidMC}}
\caption{Scalar field evolution for different values of $\beta$.} \label{phild}
\end{figure}

\begin{figure}[h!]
\centering \subfloat[][ $V(\phi_{NMC}(t))$ for different values of  $\beta$.]{
\includegraphics[scale=0.47]{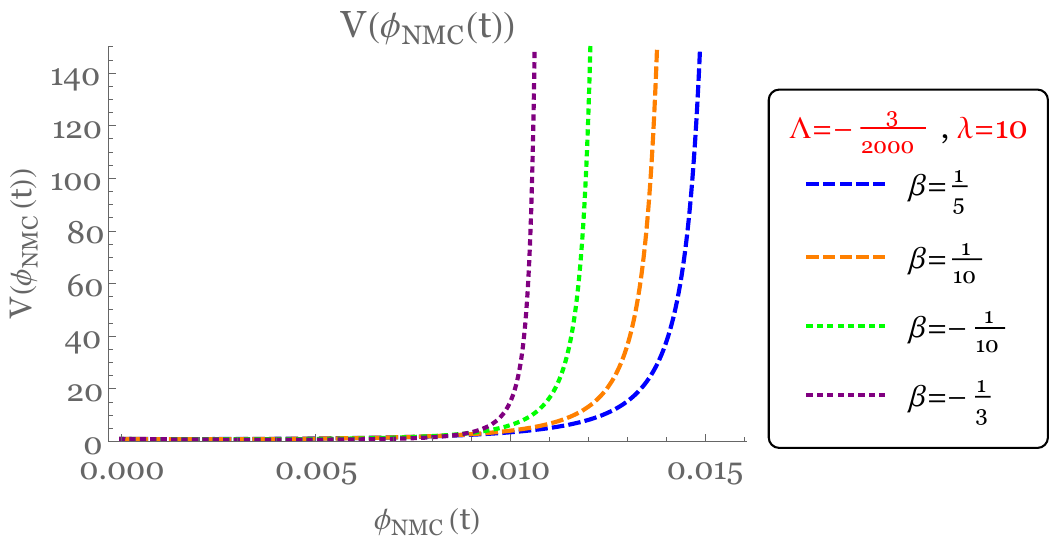}  \label{VphiNMC}} \quad \quad
\centering \subfloat[][$V(\phi_{MC}(t))$ for different values of  $\beta$.]{
\includegraphics[scale=0.47]{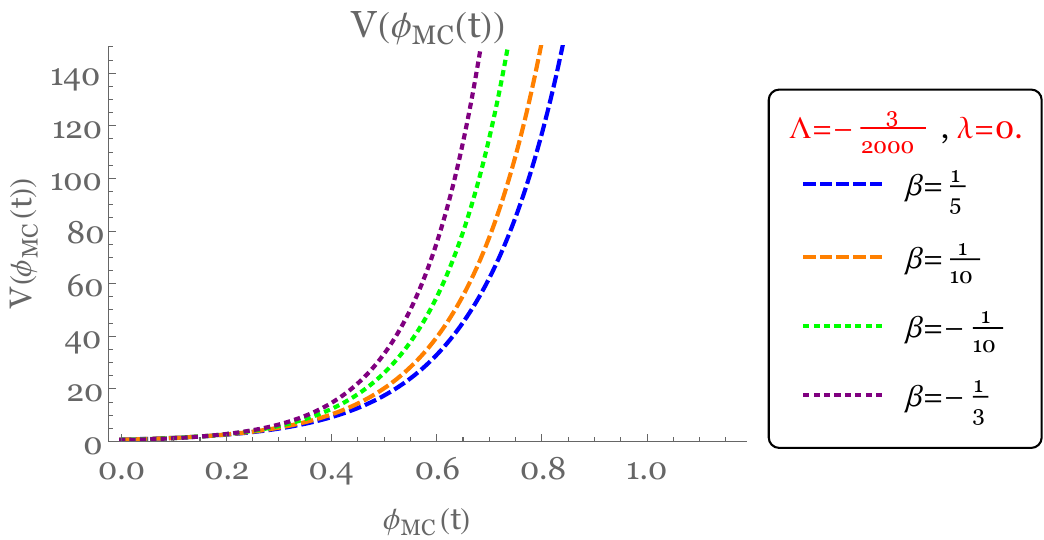}\label{VphiMC}}
\caption{Potential solution for different values of $\beta$.}
\label{Vphil}
\end{figure}

Adopting the scale factor of relation (\ref{sola}) we may proceed and solve the corresponding equations for the
scalar field and the potential.  We  differentiate  the Friedmann equation with respect to $t$ yielding
\be \label{frt}
\frac{dV(\phi(t))}{dt} = -\frac{6}{8\pi}\left(1-12\pi \lambda \dot{\phi}^2(t)\right) \left( \frac{\dot{a}^3(t)}{a^3(t)} -\frac{\dot{a}(t)\ddot{a}(t)}{a^2(t)} \right)- \left(1+ 9\lambda \frac{\dot{a}^2(t)}{a^2(t)}\right)\dot{\phi}(t) \ddot{\phi}(t)~,
\ee
and then we  multiply  the Klein-Gordon equation with $\dot{\phi}(t)$ yielding
\be  \label{kgt}
\frac{dV(\phi(t))}{dt} = -\dot{\phi}^2(t) \left(3\frac{\dot{a}(t)}{a(t)} + 3\lambda \frac{\dot{a}^3(t)}{a^3(t)} + 6\lambda \frac{\dot{a}(t)\ddot{a}(t)}{a^2(t)}\right) - \left(1 + 3\lambda\frac{\dot{a}^2(t)}{a^2(t)}\right) \dot{\phi}(t) \ddot{\phi}(t)~.
\ee
Subtracting \eqref{kgt} from \eqref{frt} we get
\be \label{eqphi}
\frac{3\dot{a}^3(t)\left(1-16\pi \lambda \dot{\phi}^2(t)\right)}{4\pi \dot{a}^3(t)} - \dot{a}(t)\left(\frac{3\dot{\phi}^2(t)}{a(t)} + \frac{3\left(1-4\pi\lambda \dot{\phi}^2(t)\right)\ddot{a}(t)}{4\pi a^2(t)}\right) + 6\lambda\frac{\dot{a}^2(t)}{a^2(t)}\dot{\phi}(t)\ddot{\phi}(t)  = 0~.
\ee
We solve \eqref{eqphi} numerically with respect to $\phi(t)$ using the expression \eqref{sola} for $a(t)$ and its derivatives,  with the initial conditions $\phi(0)=0 , \dot{\phi}(0)=0.001$, and we show the results in Fig. \ref{phil}. In the lower panel of the figure we give a the corresponding result for the minimal coupling case. As one can see, $\phi(t)$ blows up at the singularity for any  $\beta \in (-2,1)$ insofar as $\lambda = 0 $ but this is not the case for $\lambda \neq 0$. The non-minimal derivative coupling delays the evolution of the scalar field which  in turn takes a finite value at the singularity. \\
\\
We confirm this by plotting the derivative of the scalar field solution in Fig. \ref{phild}. We see once more that for $\lambda = 0$ the scalar field derivative blows up at the singularity while for $\lambda \neq 0$ it reaches a constant value.\\
\\
Finally, knowing the behaviour of the scalar field we can find the form of the potential using  \eqref{deden} and \eqref{frA}. In Fig. \ref{Vphil} we plot the potential as a function of the scalar field $\phi(t)$.

\section{Conclusions}
\label{conc}

We studied the gravitational collapse of a time-dependent scalar
field in a theory with negative cosmological constant in which,
besides its usual coupling to gravity, it is also kinematically
coupled to the Einstein tensor. No scalar field potential is present in the first stage. We checked that, during the collapsing process,
the weak energy condition is satisfied. This term is part of the Horndeski theory which gives only second-order differential equations
and is characterized by a shift symmetry.

We found that, as the
absolute value of the derivative coupling $\lambda$ is
increased,  the time $t_s$ that is needed for the scale factor to reach
the singularity also increases. Also we found that it takes more time for the apparent horizon to be formed.
The scalar field takes on a finite value at the singularity formation time $t=t_s,$ in case $\lambda$ is non-zero. This is an extra bonus, since regularity is safe and no additional actions need be taken to secure it, as has been necessary in
\cite{Kolyvaris:2011fk,Kolyvaris:2013zfa, Babichev:2013cya,Charmousis:2014zaa, Ogawa:2015pea}. Matching the metric of the collapsing scalar field with that
of an external Schwarzschild-$AdS_4$ metric we found that an event
horizon is formed that bounds the trapped apparent horizon.

To tackle the problem of the introduction of a self-interaction potential for the scalar field, we have followed a different procedure. We have used the specific ansatz $\dot{a}^2(t) = \frac{\Lambda}{3}a^2(t) + \frac{8\pi}{3}a^{2\beta}(t)$ for the scale factor $a(t),$ which has been proposed in \cite{Giambo:2005se}, \cite{Baier:2014ita}.  For negative $\beta$ we find a behavior consistent with a picture of collapse. The numerical outcome has been that an apparent horizon
appears that covers the singularity and a black hole is formed. Finally we have derived the potential $V(\phi(t))$ which is consistent with the $\beta$ value, rather than choosing it arbitrarily from the beginning.

This procedure incorporates the influence of the cosmological constant on the collapse, but has the drawback that the singularity formation time $t_s$ does not depend on the derivative coupling constant $\lambda.$ Instead it determines the profile of the scalar field and the potential in such a way that the equations of motion are satisfied with the given solution $a(t)$ corresponding to a given cosmological constant. One may consider that, if the potential has been given in advance, the relevant value of $\beta,$ hence $t_s,$ would be determined and would depend on $\lambda.$

To get a flavor of the influence of the potential on the solution, we have performed an analytic approximation valid near $t=t_s,$ for the case of the potential $V=\gamma \dot{\phi}^2,$ treated in \cite{Baier:2014ita}. We have found that $a(t) \propto (t_s-t)^{\zeta}$ along with an expression for $t_s.$
The scalar field is finite at the singularity time.
On the other hand, the analytic approximation indicates that $t_s$ may {\it decrease} with $\lambda$ for a range of positive $\gamma$ values. This is in contrast with our findings for zero potential.

It is interesting to extend the above study in the case in which the
collapsing metric and the scalar field are also space dependent. In
this case the field equations cannot be solved analytically and
numerical investigation is needed on the line of the work
presented in \cite{Garfinkle:2011hm,Garfinkle:2011tc}. This
opens up the possibility to also include in the study other terms
that appear in the Horndeski action  and it provides vital
information on the role of the Galilean symmetry in the collapsing
process. In this case we can have a direct application of the collapsing process to the thermalization of
a fluid using the gauge/gravity duality.

\acknowledgments{We benefited from discussions we had with E.
Babichev, C. Charmousis, C. Deffayet and B. Wang. G.K., K.N. and
E.P. were partially supported by ARISTEIA II action of the
operational programme education and long life learning which is
co-funded by the European Union (European Social Fund) and
National Resources. M.T was supported by the Centro de Estudios
Cient\' ificos (CECs) funded by the Chilean Government through the
Centers of Excellence Base Financing Program of the Comision Nacional de Investigacion Cientifica y Tecnologica (CONICYT), Chile. M.T. was also supported by the fellowship from the Scientific and Technological Research Council of Turkey (TUBITAK 2216) under the application number 1059B161500790.}

\begin{appendix}

\section{Analytical solution of the field equations}

In this appendix we present an exact solution of the
Klein-Gordon and the time Einstein equation.   We begin with the
Klein-Gordon equation, multiplied by
$\dot{\phi}(t),$ where we substitute $w(t)$ for the
expression $\dot{\phi}^2(t)$ \be \left(3 \lambda a(t)
\dot{a}^2(t) + a^3(t)\right) \dot{w}(t) + \left(6  \lambda
\dot{a}^{3}+ 12  \lambda a(t)
\dot{a}(t)\ddot{a}(t)+ 6 a^2(t) \dot{a}(t)\right)
w(t)=-2 \frac{dV}{dt}  a^3(t) ~, \label{KG} \ee
 where a dot denotes a derivative
with respect to time. Next we consider the  Einstein equation
performing the same substitutions \be a^2(t) [-|\Lambda| + 4 \pi w(t)] -3 \dot{a}^2(t) \left(1-12 \pi
 \lambda w(t)\right) + 8 \pi a^2(t) V=0~. \label{Einstein} \ee

We first examine the special case $V=0,\ \lambda=0.$ In that case the equations become
\be a^3(t) \dot{w}(t) + 6 a^2(t) \dot{a}(t)
w(t)=0~, \ee
\be a^2(t) [-|\Lambda| + 4 \pi w(t)] -3 \dot{a}^2(t)  =0~. \ee
Solving the second equation for $w(t)$ and substituting the
result into the first equation we get $$ \frac{\ddot{a}(t)}{a(t)} +2 \frac{\dot{a}^2(t)}{a^2(t)} +\frac{\dot{a}^4(t)}{a^4(t)}  =-|\Lambda|~. $$
Next we make the substitutions $\dot{a}(t)= H(t) a(t),\ \
\ddot{a}(t)= \frac{d H(t)}{d t} a(t) + H(t)
\dot{a}(t) = \left(\frac{d H(t)}{d t} +
H^2(t)\right)a(t).$ The result reads \be \frac{d H(t)}{d
t} = -|\Lambda|-3 H^2(t) ~.\ee Integration yields \be C_1-t = \frac{1}{\sqrt{3 |\Lambda|}} \arctan\left(\sqrt{3 |\Lambda|} H(t) \right) \Leftrightarrow H(t) = \frac{|\Lambda|}{3} \tan[\sqrt{3 |\Lambda|} (C_1-t)]~,\label{atn}\ee where $C_1$ is an integration constant. Integrating once more, equation $$ \frac{1}{a(t)} \frac{d a(t)}{d t} = \frac{|\Lambda|}{3} \tan[\sqrt{3 |\Lambda|} (C_1-t)]$$ gives \be a(t) = C_2 \left|\cos\left[\sqrt{3 |\Lambda|}(C_1-t)\right]\right|^{\frac{1}{3}},\label{exsol}\ee which subsequently gives $H(t):$ \be H(t) \equiv \frac{1}{a(t)} \frac{d a(t)}{d t} = \sqrt{\frac{ |\Lambda|}{3}}\tan[\sqrt{3 |\Lambda|}(C_1-t)]~.\ee

\section{Approximation of the solution for a specific potential near $t=t_s$}

In section IV we have postulated the ansatz (\ref{sola}) for $a(t)$ and reconstructed the potential afterwards. We now would like to take a different path and approximate the solution when the derivative coupling is present for the specific potential studied in \cite{Baier:2014ita}, which is given by the expression $$V= \gamma \dot{\phi}^2(t) \equiv \gamma w(t).$$

As a warm up we start from equation (\ref{atn}) and  set $\frac{1}{\sqrt{3 |\Lambda|}} \equiv A,\ \sqrt{3 |\Lambda|} \equiv B,$ so this equation may be rewritten in the form
\be C_1-t = A \arctan\left(B H(t) \right).\ee Since we are interested in the regime $t\to t_s,$ so that $H(t)\to -\infty,$ we just expand the right hand side about $-\infty$ and we get \be C_1-t \approx -\frac{A \pi}{2}-\frac{A}{B H(t)},\ t\to t_s~.\ee Solving we find \be H(t) = \frac{1}{a(t)} \frac{d a(t)}{d t} \approx \frac{A}{B \left(t-C_1-\frac{A \pi}{2}\right)} = -\frac{A}{B \left(t_s-t\right)},\ t_S\equiv C_1+\frac{A \pi}{2}~.\ee The expression for $a(t)$ reads \be a(t)=C_2 (t_s-t)^{\zeta},\ \zeta\equiv\frac{A}{B}~.\ee
Of course, substituting $A$ and $B$ we recover for the exponent the value $\frac{A}{B}=\frac{1}{3}.$ We see that the approximate expression incorporates the essential features of its exact counterpart, equation
(\ref{exsol}), namely it vanishes at $t=t_s$ and its derivative diverges towards $-\infty$ at the same point.

Now we try to retrace the previous steps in the case $\lambda \ne 0,\ V \ne 0~.$ It turns out that if one chooses the potential used in \cite{Baier:2014ita}, which is given by the expression $$V= \gamma \dot{\phi}^2(t) \equiv \gamma w(t) \Rightarrow \frac{dV}{dt} = \gamma \dot{w}(t)~,$$ the calculations are feasible. The above equations become
 \be \left(3 \lambda a(t)\dot{a}^2(t) + a^3(t)\right) \dot{w}(t) + \left(6  \lambda
\dot{a}^{3}+ 12  \lambda a(t) \dot{a}(t)\ddot{a}(t)+ 6 a^2(t) \dot{a}(t)\right)
w(t)=-2 \gamma \dot{w}(t)  a^3(t) ~, \label{KG1}\ee
\be a^2(t)\left (-|\Lambda| + 4 \pi w(t)\right) -3 \dot{a}^2(t) \left(1-12 \pi
 \lambda w(t)\right) + 8 \pi a^2(t) \gamma w(t)=0~. \label{Einstein1} \ee
Solving equation (\ref{Einstein1}) for $w(t)$ and substituting the result into equation (\ref{KG}) we get
$$ |\Lambda| (1+2\gamma) +\frac{\dot{a}^2(t)}{a^2(t)} \left(2+2\gamma-4 \gamma^2 +13 \lambda |\Lambda|+8 \gamma\lambda|\Lambda|\right) +\frac{\dot{a}^4(t)}{a^4(t)} \left(27 \lambda +18\lambda^2 |\Lambda|\right)
$$ $$ +27 \lambda^2 \frac{\dot{a}^6(t)}{a^6(t)} + \frac{\ddot{a}(t)}{a(t)}\left(1+4\gamma+4\gamma^2-\lambda|\Lambda|-2 \gamma\lambda|\Lambda|\right)$$ $$ + \frac{\ddot{a}(t)}{a(t)}\frac{\dot{a}^2(t)}{a^2(t)}\left(9\lambda+18\gamma\lambda+9 \lambda^2|\Lambda|\right) + 54\lambda^2 \frac{\ddot{a}(t)}{a(t)}\frac{\dot{a}^4(t)}{a^4(t)}=0~. $$
Next we make the substitutions $\dot{a}(t)= H(t) a(t),\ \
\ddot{a}(t)= \frac{d H(t)}{d t} a(t) + H(t)
\dot{a}(t) = \left(\frac{d H(t)}{d t} +
H^2(t)\right)a(t).$ The result reads \be \frac{d H(t)}{d
t} = -\frac{[ |\Lambda|+3 H^2(t)] [1+3
\lambda H^2(t)] [1+2\gamma+9\lambda H^2(t)]}{(1+2\gamma)(1+2\gamma-
\lambda |\Lambda|) + 9 \lambda (1+2\gamma +\lambda |\Lambda|)H^2(t)+ 54\lambda^2 H^4(t)}~.\label{noV}\ee
This equation is separable and may be solved readily decomposing into partial fractions
\be \label{allin} C_1-t = A_1\arctan(B_1 H(t))  + A_2  \arctan(B_2 H(t)) + A_3 \arctan(B_3 H(t))~,\ee where
$$  A_1 = \sqrt{\frac{\lambda}{3}}\frac{2-\gamma+2 \gamma^2-2 \lambda|\Lambda|-\gamma\lambda|\Lambda|}{(\gamma-1)(\lambda |\Lambda|-1)},\ B_1=\sqrt{3 \lambda}, $$ $$ A_2= \frac{\sqrt{\lambda(1+2 \gamma)}}{\gamma-1},\ B_2= 3 \sqrt{\frac{\lambda}{1+2\gamma}},\ A_3=\frac{1+2\gamma-\lambda|\Lambda|}{\sqrt{3 |\Lambda|}(1-\lambda|\Lambda|)},\ B_3=\sqrt{\frac{3}{|\Lambda|}} ~,$$
One may set $\lambda=0,$ but one may not set the cosmological
constant to zero, since it appears in the denominators.

Taylor expanding (\ref{allin}) around infinite $H$ one gets \be C_1 -t \approx -\frac{\pi}{2}(A_1+A_2+A_3) -\frac{\frac{A_1}{B_1}+\frac{A_2}{B_2}+\frac{A_3}{B_3}}{H(t)} \Rightarrow H(t) = \frac{-\frac{A_1}{B_1}-\frac{A_2}{B_2}-\frac{A_3}{B_3}}{C_1+\frac{\pi}{2}(A_1+A_2+A_3)-t}~.\ee
We may now integrate once more using the initial condition $a(0)=1.$ Using the notations $$\zeta\equiv \frac{A_1}{B_1}+\frac{A_2}{B_2}+\frac{A_3}{B_3},\ t_s \equiv C+\frac{\pi}{2}(A_1+A_2+A_3)$$ one finds
\be H(t) = \frac{\dot{a}(t)}{a(t)} =-\frac{\zeta}{t_s-t} \Rightarrow
a(t) = C_2 \exp\left[-\int d t \frac{\zeta}{t_s-t}  \right] = C_2 \left(t_s-t\right)^\zeta.\ee In our case the parameter turns out to equal $\zeta = \frac{2}{3},$ so, finally
\be \label{approxl} a(t)= C_2 \left(t_s-t\right)^{\frac{2}{3}}\Rightarrow H(t) = -\frac{2 C_2}{3}\left(t_s-t\right)^{-\frac{1}{3}}.
\ee
This means that, for $t=t_s,$ the scale factor $a(t)$ vanishes, while $H(t)$ diverges to $-\infty,$ so the quantity $t_s$ is the singularity formation time.
The apparent horizon covers the singularity, so a black hole is formed. We note that the result is independent of the coupling $\lambda$ and the potential magnitude $\gamma!$ This is the reason why we can also claim that \eqref{approxl} is the approximate solution in section \ref{apprh} where $\gamma = 0$ with $\lambda \neq 0$.

If we adopt the expression $t_s =\frac{\pi}{2}(A_1+A_2+A_3)$ and examine the $\Lambda,\ \lambda$ and $\gamma$ dependence of $t_s$ we find qualitative agreement with the numerical results presented in the text.

Solving equation (\ref{Einstein1}) we find the expression: \be w(t) = \frac{1}{4 \pi} \frac{3 \dot{a}^2(t)+|\Lambda| a^2(t)}{9 \lambda \dot{a}^2(t) +(1+2 \gamma)  a^2(t)}~.\ee If we assume that $\lambda\ne 0$ and  substitute the solution for $a(t)$ just found, we end up with the constant $\frac{1}{12 \pi \lambda}$ as the limit of $w(\tau)=\dot{\phi}^2(t)$ as $t\to t_s.$

On the other hand, if $\lambda = 0,$ that is if the derivative coupling is absent, the remaining expression \be w(t) = \frac{1}{4 \pi} \frac{3 \dot{a}^2(t)+|\Lambda| a^2(t)}{(1+2 \gamma)  a^2(t)}\ee yields a scalar field that diverges on the horizon. This result agrees with the results of \cite{Baier:2014ita} and shows that the presence of the non-minimal coupling changes drastically the behaviour of the system.

In \cite{Baier:2014ita} the ansatz $\dot{a}^2(t) = a^{2 \beta}(t) -\frac{1}{l^2}$ has been used, where $l$ is the radius corresponding to the cosmological constant. In the limit $t \to t_s$ the ansatz may be written in the form $\dot{a}(t) \approx -a^{\beta}(t) \Rightarrow H(t) \approx -a^{\beta-1}(t).$ $\beta$ has been assumed to take on negative values, so that $\dot{a}(t) $ tends to $-\infty$ and dominates.
Substitution of the ansatz into (\ref{noV}) gives $$ 81 \lambda^2 a^{6 \beta-6}(t)+9\lambda a^{4 \beta-4}(t)(4+2\gamma+3 \lambda |\Lambda|+6 \lambda (\beta-1)a^{2 \beta-2}(t)) +(1+2\gamma)(|\Lambda|+(\beta-1)(1+2\gamma-\lambda|\Lambda|) a^{2 \beta-2}(t)$$ $$+3 a^{2 \beta-2}(t)(1+2\gamma+4 \lambda|\Lambda| +2 \gamma\lambda|\Lambda| - 3 \lambda (\beta-1)(1+2\gamma+\lambda|\Lambda|)a^{2 \beta-2}(t)=0~.$$

In our case, where $\lambda \ne 0,$ we may keep the dominant contributions, which are  the terms multiplying $a^{6 \beta-6}(t)$ and get
$$81 \lambda^2 + 54 \lambda^2 (\beta-1)=0 \Rightarrow \beta=-\frac{1}{2}~.$$ That is, once $\lambda$ is introduced, the value of $\beta$ is fixed. The ansatz becomes $\dot{a}(t)=-a^{-\frac{1}{2}}(t),$ which yields the solution $a(t)= C_2 \left(t_s-t\right)^{\frac{2}{3}}$ found before.

In \cite{Baier:2014ita} the potential $V= \gamma \dot{\phi}^2(t) \equiv \gamma w(t)$ has been examined in the case $\lambda=0.$ We substitute this ansatz into equation (\ref{noV})  for $\lambda=0$ and isolate the dominant terms, which are the ones multiplying $a^{2 \beta-2}(t),$ yielding $$3(1+2\gamma)+(1+2\gamma)((\beta-1)(1+2\gamma)=0\Rightarrow \gamma = \frac{\beta+2}{2(1-\beta)},$$ which is exactly the relation  given in \cite{Baier:2014ita}.

Thus, with this specific form of the potential, switching on the non-minimal coupling restricts $\beta$ to a particular value. Other forms of the potential may allow for other values of $\beta.$

We remark that $t_s$ increases with $\lambda$ for $\gamma=0.$ However, for different values of $\gamma$ things may be different.
\begin{figure}[ht!]
\begin{center}
\includegraphics[scale=0.5]{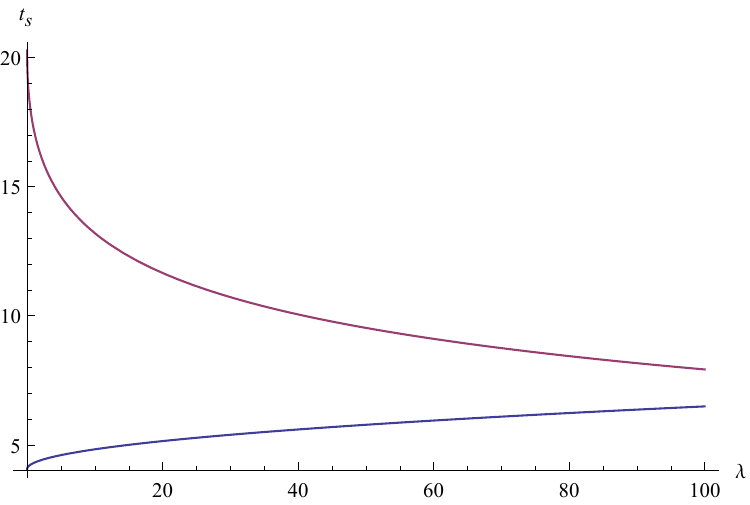}
\end{center}
\caption{Singularity formation time $t_s$ versus $\lambda$ for $\gamma=0$ (the increasing curve) and for $\gamma=2.0$ (the decreasing curve). $|\Lambda|=0.05$ for both curves.}
\label{gamma}
\end{figure}
In Fig. \ref{gamma} we show two plots of  $t_s$ versus $\lambda$ for $|\Lambda|=0.05,$ one for $\gamma=0$ and one for $\gamma=2.0.$ We see that the behaviours are different.

\end{appendix}


\end{document}